\documentclass[sigconf]{acmart}

\setcopyright{none}
\renewcommand\footnotetextcopyrightpermission[1]{}
\acmConference{}{}{}

\usepackage[english]{babel}
\usepackage{balance} 
\usepackage{dblfloatfix}
\usepackage{float}
\usepackage{enumitem} 
\usepackage{graphicx}
\usepackage{caption}
\usepackage{subcaption}
\usepackage{multirow} 
\usepackage{appendix} 

\PassOptionsToPackage{hyphens}{url}\usepackage{hyperref}
\begin{document}
\title{Topology and Geometry of the Third-Party Domains Ecosystem: Measurement and Applications}

\author{Costas Iordanou } 
\affiliation{%
  \institution{Cyprus University of Technology}
  \country{}
  }
\email{kostas.iordanou@cut.ac.cy}

\author{Fragkiskos Papadopoulos}
\affiliation{%
  \institution{Cyprus University of Technology}
  \country{}  
}
\email{f.papadopoulos@cut.ac.cy}

\begin{abstract}
Over the years, web content has evolved from simple text and static images hosted on a single server to a complex, interactive and multimedia-rich content hosted on different servers. As a result, a modern website during its loading time fetches content not only from its owner's domain but also from a range of third-party domains providing additional functionalities and services. Here, we infer the network of the third-party domains by observing the domains' interactions within users' browsers from all over the globe. We find that this network possesses structural properties commonly found in complex networks, such as power-law degree distribution, strong clustering, and small-world property. These properties imply that a hyperbolic geometry underlies the ecosystem's topology. We use statistical inference methods to find the domains' coordinates in this geometry, which abstract how popular and similar the domains are. The hyperbolic map we obtain is meaningful, revealing the large-scale organization of the ecosystem. Furthermore, we show that it possesses predictive power, providing us the likelihood that third-party domains are co-hosted; belong to the same legal entity; or merge under the same entity in the future in terms of company acquisition. We also find that complementarity instead of similarity is the dominant force driving future domains' merging. These results provide a new perspective on understanding the ecosystem's organization and performing related inferences and predictions.
\end{abstract}

\ccsdesc[300]{General and reference~Measurement}
\ccsdesc[500]{Networks~Network domains}
\ccsdesc[500]{Networks~Network measurement}
\ccsdesc[300]{Networks~World Wide Web (network structure)}
\ccsdesc[300]{Theory of computation~Graph algorithms analysis}
\ccsdesc[500]{Theory of computation~Computational geometry}
\ccsdesc[500]{Mathematics of computing~Geometric topology}

\keywords{Third-Party Domains, Network Topology, Hyperbolic Embedding, Applications}

\maketitle

\section{Introduction}
Over the years, web content has evolved from simple text and static images hosted on a single server to a complex, interactive and multimedia-rich content hosted on different servers. As a result, a modern website during its loading time fetches content not only from its owner's domain (first-party domain) but also from a range of third-party domains (TPDs) that provide additional functionalities, such as advertising agencies, content distribution networks, and analytics services.

The increasing complexity of websites and the resulting TPD ecosystem have been thoroughly studied from different perspectives. A plethora of work focuses on privacy issues arising from third-party tracking~\cite{Acar2013, englehardt2016, englehardt2015, Gomer2013, Schelter2016, Takano2014, Falahrastegar2014, Quan2021, Bashir2016, Solomos2019, Urban2020} and web advertising~\cite{Iordanou2019, Carrascosa_2015, Bashir2018, Pujol_2015, Walls_2015, Bangera2017, Venkatadri_2018}, and on the impact of different privacy protection regulations on them~\cite{Iordanou_2018, Reyes2018, Razaghpanah2018}, such as the General Data Protection Regulation in Europe (GDPR)~\cite{gdpr}, the California Consumer Privacy Act~\cite{CCPA}, and the Children's Online Privacy Protection Rule in the US~\cite{coppa}. The literature also includes studies related to malvertising~\cite{Shubham2017, Muazzam2008, Aditya2011}, crypto-miners~\cite{Konoth2018, Jan2018, Shayan2018}, user experience~\cite{Butkiewicz2011, Brajnik2010, Dhote2013}, etc. However, in all existing studies the analysis focuses only on a subset of TPDs that are relevant to the corresponding type of analysis (e.g., web tracking and advertising domains in order to study the online advertising ecosystem).

Differently from existing work, here we attempt to provide for the first time a holistic characterization of the TPD ecosystem as a large-scale network. In this network, a connection between two TPDs signifies that these TPDs have interacted within a user's browser. To capture the interactions between TPDs we monitor HTTP(s) requests within users' browsers from all over the world and consider all TPDs, irrespective of their activities.\footnote{The ``s'' in HTTP(s) is to indicate that we also track secure HTTP requests. For brevity, in the rest of the paper we drop the ``s''.} We find that the resulting network possesses structural properties commonly found in complex networks, such as power-law degree distribution, strong clustering, and small-world property~\cite{Dorogovtsev10-book}. These properties in turn imply that a hyperbolic geometry underlies the ecosystem's topology~\cite{Krioukov2010} and we use statistical inference methods to infer the TPDs coordinates in this geometry~\cite{Garc_a_P_rez_2019}. 

The inferred TPD coordinates abstract how popular and similar~\cite{Papadopoulos2012} the TPDs are, while the hyperbolic map we obtain, which we visualize, is remarkably meaningful. The map showcases the large-scale organization of the TPD ecosystem, and reveals interactions between controversial services and social networks that have not been previously revealed. Furthermore, we show that the map possesses predictive power, providing us the likelihood that TPDs are co-hosted; belong to the same legal entity; or merge under the same entity in the future in terms of company acquisition. The merging of TPDs can happen for different reasons as we explain in Sect.~\ref{sec:applications}. However, we find that the majority of mergings are among TPDs providing complementary instead of similar functionalities. We are interested in TPD co-hosting and merging, as these phenomena may imply collaboration between the involved TPDs~\cite{postGDPR, openRTB, srinivasan2020google, Doh-Shin2021}. Thus, related inferences and predictions could be useful for regulators and data protection authorities in conducting investigations related to TPDs that may be collaborating in web tracking, fraudulent activities or misbehavior.

In summary, our main findings in this paper are the following:
\begin{enumerate}[leftmargin=0.5cm]
    \item[(1)] the TPD network has the typical properties of complex networks (power-law degree distribution, strong clustering, small-world property, etc.~\cite{Dorogovtsev10-book});
    \item[(2)] the TPD network admits a meaningful hyperbolic embedding, which allows us to readily visualize communities of TPDs and their interconnections; in this visualization, we observe connections among TPDs suggesting interactions related to web advertising and tracking, as well as close connections between controversial services (related to cryptocurrency ads and Pay-To-Click advertising services) with social networks;
    \item[(3)] hyperbolically closer TPDs have higher chances of being co-hosted; grouped currently under the same legal entity; or merge in the future under the same entity;
    \item[(4)] complementarity instead of similarity is the dominant force driving future TPD mergings.
\end{enumerate}
Additionally, we find that:
\begin{enumerate}[leftmargin=0.5cm]
    \item[(5)] interaction chains between the TPDs, as observed in the wild from real users' browsers, follow almost always shortest paths in the TPD network;
    \item[(6)] the TPD network is highly navigable~\cite{Boguna2007}, in the sense that a simple greedy routing strategy that uses the hyperbolic coordinates of the TPDs can find intended communication targets with high probability, following almost shortest paths;
    \item[(7)] TPD networks, as observed across five different geographic continents (Europe, South America, Asia, Africa and North America), have qualitatively similar topological characteristics as that of the global TPD network and exhibit high percentages of common nodes and links.
\end{enumerate}

The rest of the paper is organized as follows. In Sect.~\ref{sec:dataset} we describe the data collection process, our data pre-processing steps, and the different metadata that we use. Section~\ref{sec:graph} describes the construction of TPD networks at two different aggregation levels and analyzes their topological properties. Section~\ref{sec:hyperbolic} describes the hyperbolic embedding process and visualizes the resulting hyperbolic map of the TPD ecosystem. In Sect.~\ref{sec:applications} we show that the hyperbolic distance between TPDs provides information about their merging and co-hosting probabilities. In Sect.~\ref{sec:communication} we analyze the interaction paths between TPDs and study the navigability of the ecosystem. Section~\ref{sec:related} discusses related work and we conclude in Sect.~\ref{sec:conclusions}.

Together with this paper we release the following artifacts:
\begin{itemize}[leftmargin=0.5cm]
    \item[(a)] the raw dataset used in this work; we note that this dataset contains additional information, beyond of what used in this work, which could allow further analysis and study of the TPD ecosystem;
    \item[(b)] the different metadata that we use;
    \item[(c)] the tools that we have developed in order to collect the different metadata that we use, as well as our hyperbolic visualization tool;
    \item[(d)] the source code used to compute our results from the raw data.
\end{itemize}
More information on the released artifacts is provided in the sections that follow and in Appendix~\ref{appendix}. The artifacts are self-contained and available at~\cite{TPD_repo}.

\section{Terminology and Datasets} 
\label{sec:dataset}

\subsection{Terminology} 
\label{subsec:preliminaries}
We first provide some information related to the terminology we use in the rest of the paper.
\\
\noindent\textbf{First-party domain (FPD).} Let us assume that a user types in the browser's address bar the following URL, ``\url{https://www.example.com/about}''. In this case, the ``www.example.com'' is a first-party domain.
\\
\noindent\textbf{Third-party domain (TPD).} Following our example above, let us assume that we are able to monitor the user's browser and observe all HTTP requests that the visited website ``\url{https://www.example.com/about}'' is triggering in order to be fully loaded at the user's browser. A third-party domain is any domain in an HTTP request's URL that is different from ``www.example.com''. For instance, if we observe an HTTP request towards the URL ``\url{https://www.facebook.com/like}'', ``www.facebook.com'' is a third-party domain.
\\
\noindent\textbf{TPD Interactions.} While a website is rendering we consider any invoked HTTP request between two different TPDs (at any direction) a TPD interaction.
\\
\noindent\textbf{TPD co-hosting.} We consider two TPDs to be co-hosted if they use the same IP address (see Sect.~\ref{subsec:data_processing} for details). This can be deduced by inspecting the header of the HTTP responses from the TPDs.    
\\
\noindent\textbf{Legal entities.} Different TPDs are owned by different companies which in turn may belong to different organizations. We refer to each such organization as the legal entity that groups together all TPDs from the companies that fall under the organization.
\\
\noindent\textbf{Domain names and aggregation levels.} Let us consider the following URL ``\url{https://www.mail.example.com/about}''. The \textit{fully qualified domain name (FQDN)} in this example is ``www.mail.example.com''. The \textit{top level domain (TLD)} is ``.com'', the \textit{TLD+1} is ``example.com'', ``mail'' is a \textit{sub-domain}, ``www'' is the \textit{hostname} and ``/about'' is the \textit{path}.

\subsection{Data Collection Challenges} \label{subsec:Challenges}
Any study related to TPD tracking needs to address two main challenges. The first challenge is related to the geographic diversity of the TPD ecosystem. Specifically, the same website can trigger requests to different TPDs depending from where geographically the website is called~\cite{Falahrastegar2014, Schelter2016, Falahrastegar2016, Iordanou_2018}. The second challenge is related to the dynamic triggering of additional TPDs from the users' interaction(s) with a website's content~\cite{Urban2020, Iordanou_2018, Bashir2016, Bashir2018}. Thus, to maximize the number of observed TPDs one needs to collect website visits from real users located at different geographic regions. Using real users as opposed to web crawling techniques has the additional benefit of capturing the interactions of the users with the websites' contents, and thus the requests triggered to additional TPDs.

Given the above considerations, we collect TPD data using a browser extension that we have developed for the Google Chrome browser. This extension allows real users to interact with websites from different geographic regions from all over the world. We developed this browser extension to study targeted advertising in a prior work~\cite{Iordanou2019}. Here, we use TPD interaction data we collected using this extension from October 2017 to March 2020. It is based on these data that we construct the TPD network, as we describe in the next section. More details on the data collection process are given below.

\subsection{Data collection process} \label{subsec:data_collection}
During the rendering time of a website the first-party domain embeds a JavaScript code in order to invoke different TPDs that provide additional functionalities and services. The invoked TPDs may in turn embed their own code in order to invoke other TPDs that they collaborate with, and so on, thus creating a chain of interactions. Using our browser extension and by utilizing different browser extension APIs, such as the webRequests~\cite{chromeWebRequests} and webNavigation~\cite{chromeWebNavigation} APIs, we are able to monitor and capture the interactions between the different TPDs during the rendering time of a website.

User recruitment is always a challenge in all studies that depend on real users. To deal with this issue we performed the following. First, we released our browser extension to the public through the Google Chrome web store~\cite{chrome_webstore}. Second, to motivate users to install our extension we spread the word in online social networks. Third, we encouraged friends and colleagues around the world to install our extension. Finally, to bootstrap our user base, we also recruited users from the Figure Eight platform (currently acquired by appen~\cite{appen}). The recruited users (50) contributed data in a time window of four months, from October 2017 to January 2018, while the rest of the users (431) from the Google Chrome web store continued to contribute data until March 2020, where we ended the data collection process. Overall, we collected data from 481 users from 56 countries across the globe. The data includes visits to more than 113K websites belonging to more than 7K domains. The total number of HTTP requests that we observed is more than 11M, from which 3.4M belong to interactions between TPDs. The total number of observed TPDs is more than 10K at the FQDN level, which corresponds to $1847$ TPDs at the TLD+1 level. We publicly release the above data, see Appendix~\ref{app:raw} for further details. 

\noindent\textbf{Ethical considerations:} The browser extension complies with all requirements of the Google Chrome web store including the protection of users' privacy. We also clearly explain in the privacy policy of the extension what data the extension is collecting and for what reason. In addition, we refrained from asking or collecting any personally identifiable information such as the real name of the users, emails, addresses, etc. We note that our browser extension activates the data collection process only if the user provides explicit consent after reading our user consent statement that clearly explains what the extension is actually collecting and when. With respect to the different datasets that we release to the public, all of them are related to TPD interactions and the legal entities the TPDs belong to. No personal data are included or need to be anonymized.

\subsection{Data pre-processing and metadata} \label{subsec:data_processing}
Below we provide details related to our data pre-processing steps and the metadata that we use.  
\\
\noindent\textbf{1. TPDs detection.} Since our goal is to identify all TPDs without considering the type of service they provide, our detection process is relatively simple compared to prior studies that were focusing on a specific type of TPDs (e.g., web-tracking-related TPDs~\cite{Iordanou_2018, Iordanou2019, Razaghpanah2018,Gervais2017, Bangera2017, englehardt2016, Bashir_sockets, Leung_2016, Binns_2018, Lerner2016, Walls_2015, Pujol_2015, Quan2021}). Here, if we observe a request towards a domain other than the one that the user is actually visiting, we consider that domain a TPD.
\\
\noindent\textbf{2. TPD co-hosting.} Apart from detecting TPDs, the browser extension is also able to extract the IP address of the corresponding TPD servers. To this end, the extension monitors all HTTP requests and extracts the IP addresses from the response headers. We consider that TPDs are co-hosted if they share the same IP address. This co-hosting may be on the same physical server or under the same data center. We note that we consider IPs that are shared only by TPDs. To ensure this, we apply the methodology proposed in~\cite{Iordanou_2018} that is based on passive DNS replication (pDNS)~\cite{pdns}. We note that using the reverse DNS records, we can identify all the domains that operate on a specific IP address within our data collection period (October 2017 to March 2020).

Specifically, we observe 1181 pairs of TPDs that share the same IP address. We call the list of these 1181 pairs the \textit{``co-hosting list''}. For each TPDs in this list we also record the legal entity where it belongs to (see below).
\\
\noindent\textbf{3. From FQDNs to TLD+1.} As a next step, we choose to group together the FQDNs of the observed TPDs that belong to the same TLD+1 level. This choice is based on the following observations. First, we consider how the Same Origin Policy~\cite{SOP} works, that is, the TLD+1 domain level can read all web cookies from all $\text{TLD}+N$, $N=2,3,\dots,M$, where $M$ is the maximum number of sub-domains that a TLD+1 has. For instance, the cookies owned by \textit{a.example.com} and \textit{b.example.com} can be accessed by \textit{example.com}. Second, web cookies are the main way of tracking users on the web, exchanging information and identifiers across domains~\cite{Quan2021, Iordanou_2018, Bashir2018}. Given these observations, the two domains \textit{a.example.com} and \textit{b.example.com} can be considered ``direct collaborators'' and can be aggregated under the single domain of their corresponding TLD+1 level, \textit{example.com}. After applying this step we end up with $1847$ TLD+1 domains. 
\\
\noindent\textbf{4. From TLD+1 to legal entities.} Most domains belonging to the same legal entity share information about their users. For example, Facebook collects data from WhatsApp and Instagram~\cite{facebook_share}. Therefore, we also consider a coarser-grained aggregation of the TPD ecosystem, where we group together the TLD+1 that belong to the same legal entity, similar to other studies~\cite{Quan2021, Karaj2018, Bashir2018}. For instance, the ``Google'' legal entity groups together all TLD+1 that belong to Google (e.g., youtube.com, google.com, doubleclick.com, etc.). To identify the TLD+1 that belong to the same entity and the type of their online activities we use data from three different sources: (1) whotracks.me~\cite{whotracks_me}; (2) disconnect.me~\cite{disconnect_me}; and (3) the webXray domain owners list~\cite{webXray}. For the TLD+1 not included in these sources, we manually examine them using the crunchbase website~\cite{crunchbase} in order to extract information related to the legal entity behind them and the type of their online activities. Overall we had to make this manual inspection for 243 TLD+1 domains. We refer to this dataset as the \textit{``legal-entity list''}.

\noindent \textbf{5. Future TLD+1 mergings.} We also use the crunchbase website to collect information related to TLD+1 acquisitions by legal entities after the end of the data collection process of Sect.~\ref{subsec:data_collection}, that is, from April 2020 up to November 2021. To facilitate this process we developed a tool that we make publicly available (see Appendix~\ref{appendix}). We call these acquisitions \emph{future TLD+1 mergings} and refer to this dataset as the \textit{``future-merging list''}.




\section{The TLD+1 and Legal-entity Networks} \label{sec:graph}

We consider the following two networks that correspond respectively to a finer- and a coarser-grained aggregation of TPDs.

\begin{figure}[t]
\centering
\includegraphics[width=0.9\columnwidth]{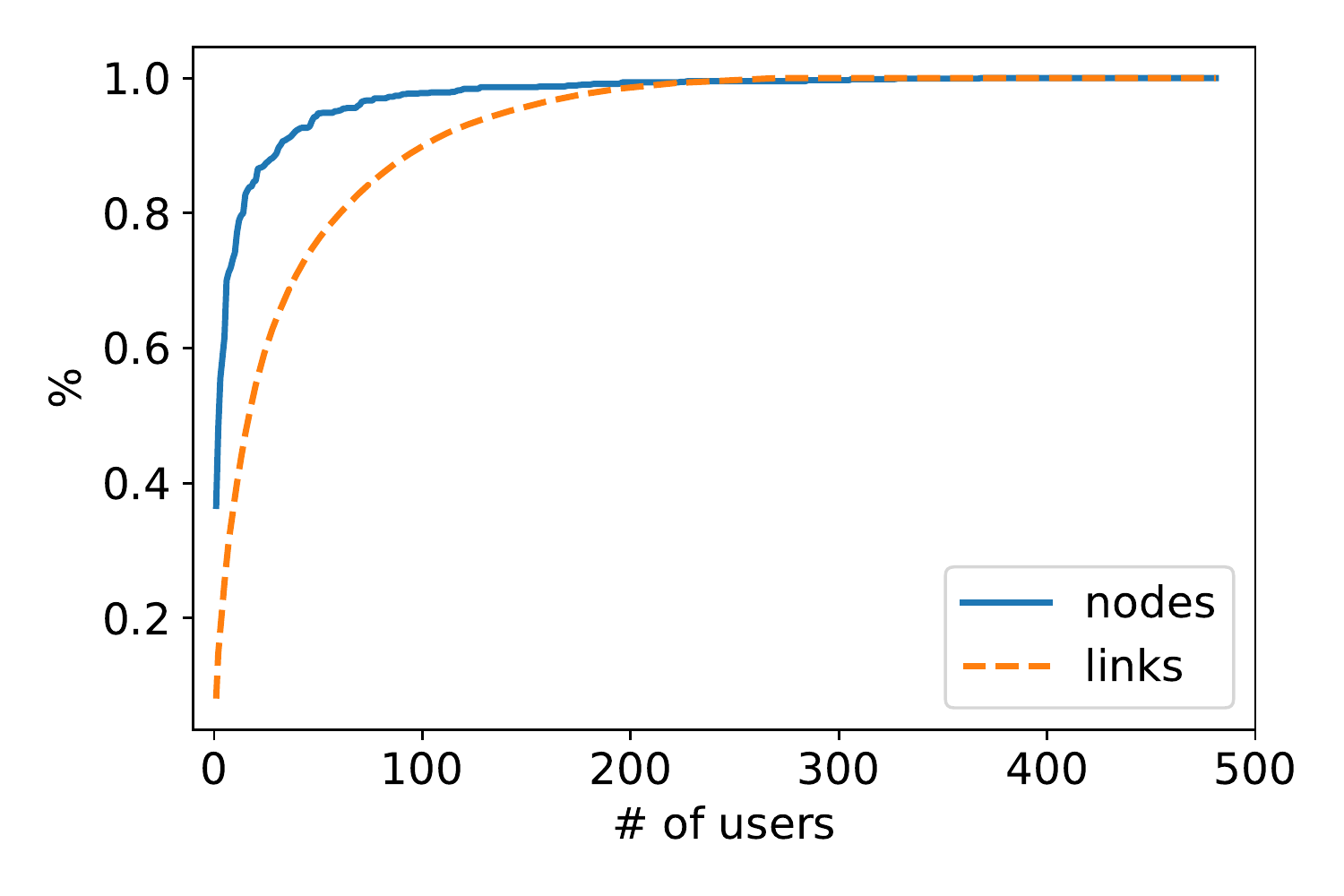}
\vspace{-6mm}
\caption{Percentage (out of the total) of discovered TLD+1 nodes (solid line) and discovered TLD+1 links (dashed line) introduced to our TLD+1 network as a function of the number of users (see text).}
\label{fig:NodesPerUser}
\end{figure}

\noindent\textbf{The TLD+1 network.} In this network nodes are TPDs aggregated at the TLD+1 level. There is an edge between two TLD+1 domains if they interact at least once within a user's browser. In Sect.~\ref{sec:merging} we show that the hyperbolic embedding of this network provides information about the likelihood of current grouping or future merging of TLD+1 domains into legal entities.

\noindent\textbf{The legal-entity network.} In this network nodes are legal entities. There is an edge between two legal entities if there is at least one edge between the TLD+1 domains that belong to the entities. In Sect.~\ref{sec:visualization} we visualize the hyperbolic embedding of this network, which shows the different communities of the TPD ecosystem. Furthermore, in Sect.~\ref{sec:co-hosting} we show that the hyperbolic embedding of this network provides information about the co-hosting likelihood of legal entities.

Both networks are connected. An overview of their basic topological properties is given in Table~\ref{tab:networkGraphsStats}. We see that both networks are sparse ($\bar{k} \ll N$) and small-worlds ($d_\textnormal{max} \ll N$). They also have high clustering ($\bar{c}$) and heterogeneous degrees ($k_\textnormal{max} \gg \bar{k}$). These characteristics are typical of complex networks found in many other domains~\cite{Dorogovtsev10-book}. 

\begin{figure*}[ht]
     \centering
     \hfill
     \begin{subfigure}[b]{0.33\textwidth}
         \centering
         \includegraphics[width=\textwidth]{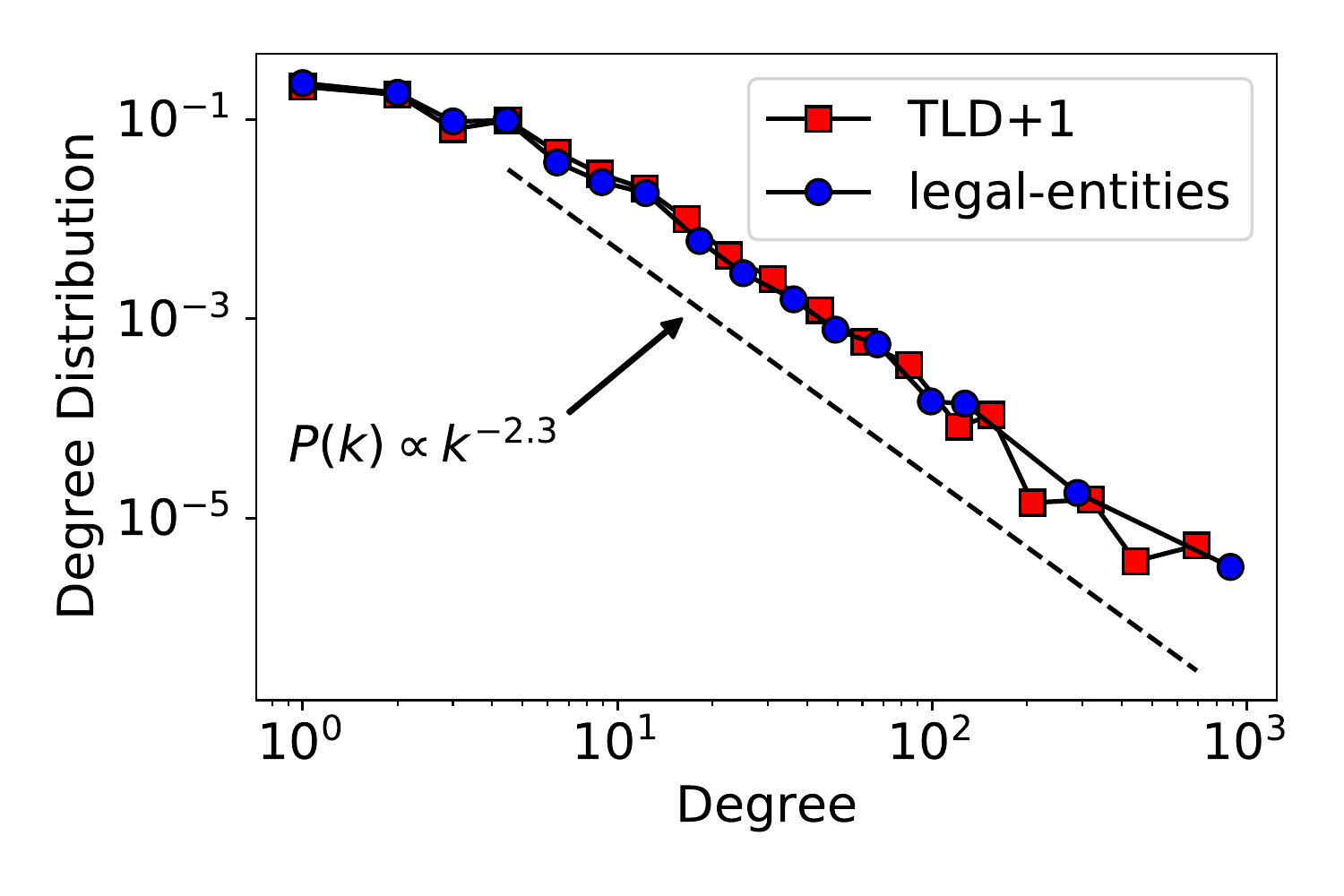}
         \vspace{-7mm}
         \caption{Degree distribution $P(k)$}
         \label{subfig:degree_distributions}
     \end{subfigure}
     \hfill
     \begin{subfigure}[b]{0.33\textwidth}
         \centering
         \includegraphics[width=\textwidth]{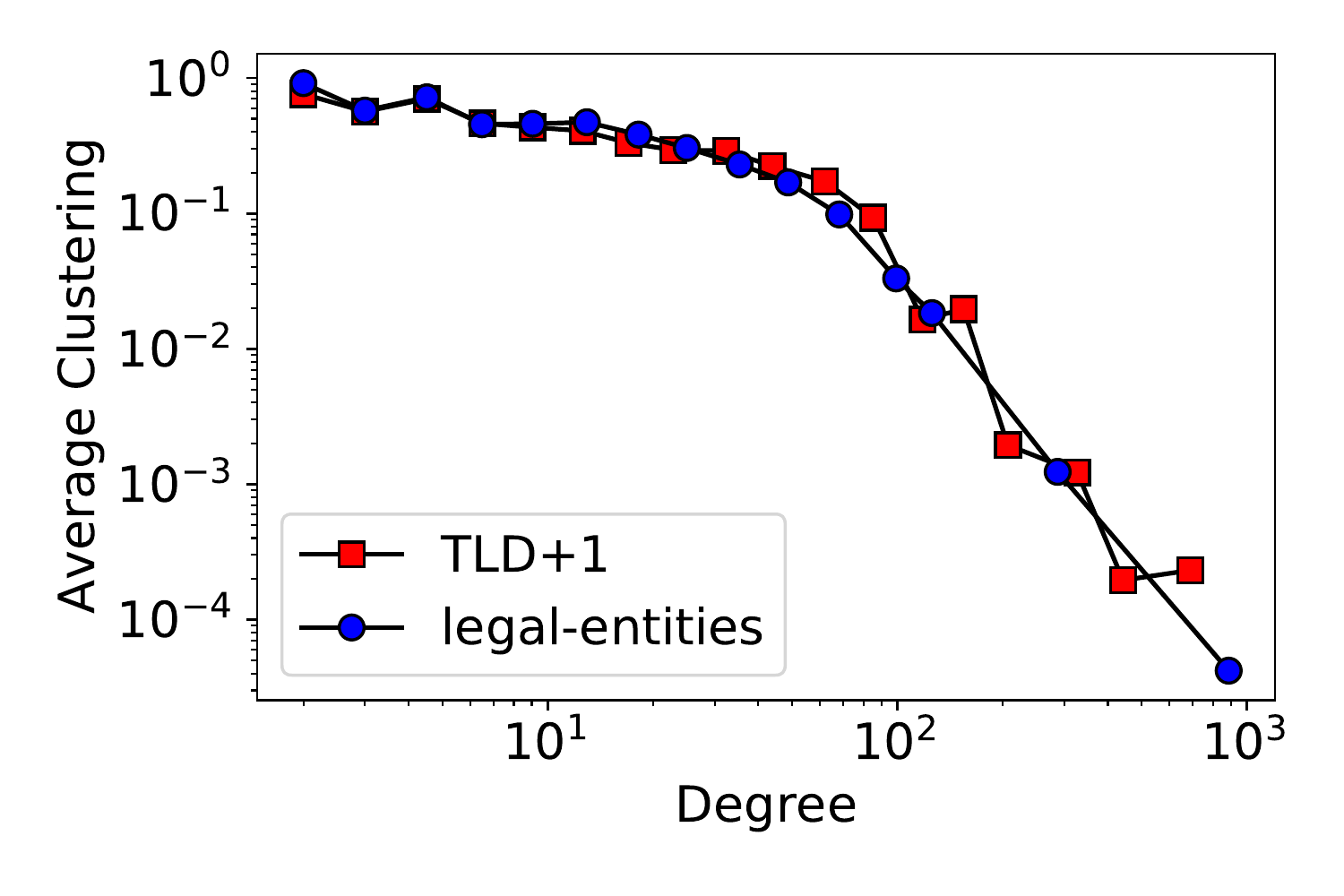}
         \vspace{-7mm}
         \caption{Average clustering $\bar{c}(k)$}
         \label{subfig:average-clustering}
     \end{subfigure}
     \hfill
     \begin{subfigure}[b]{0.33\textwidth}
         \centering
         \includegraphics[width=\textwidth]{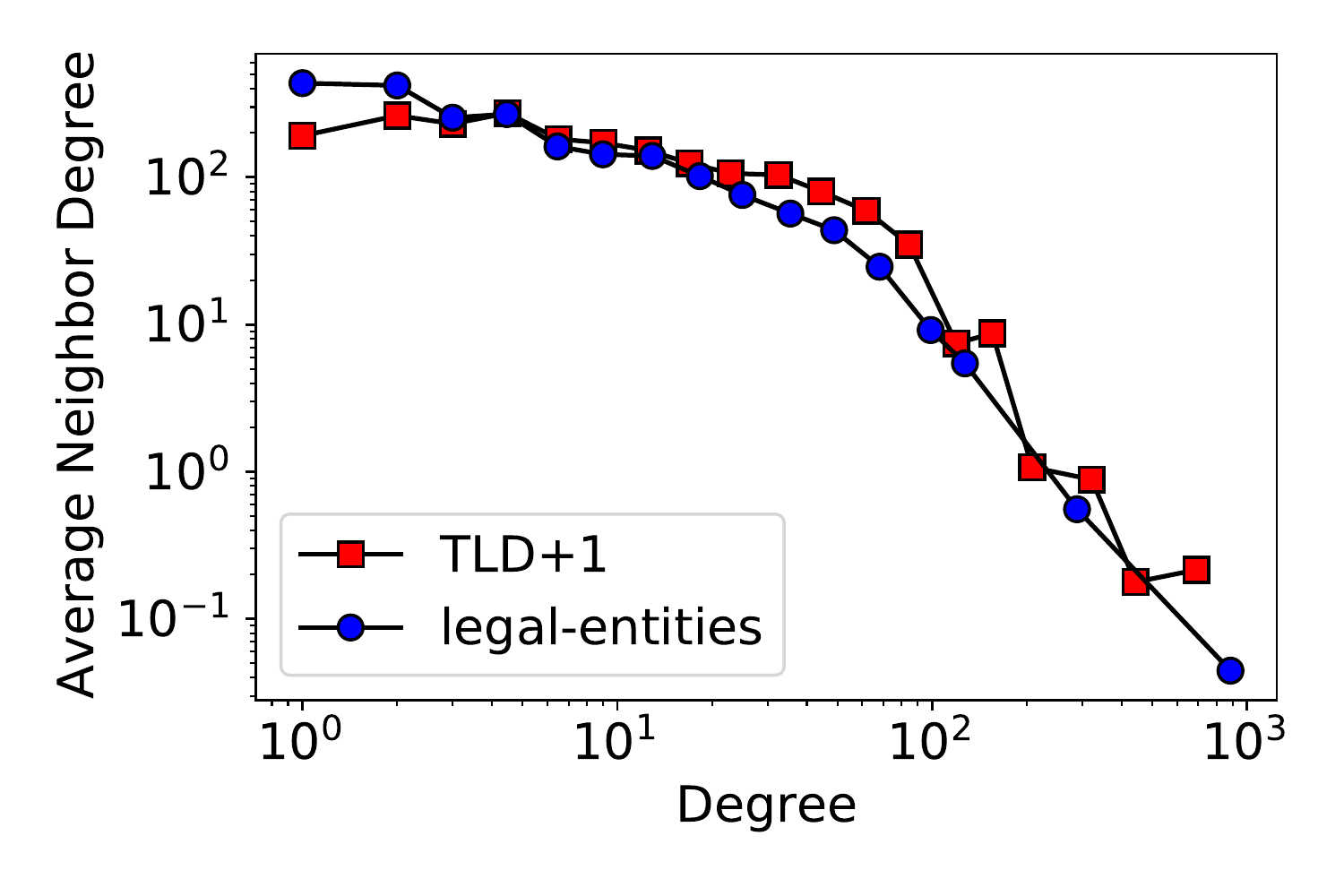}
         \vspace{-7mm}
         \caption{Average neighbor degree $\bar{k}_{\text{nn}}(k)$}
         \label{subfig:average-neighbor-degree}
     \end{subfigure}
     \begin{subfigure}[b]{0.31\textwidth}
         \centering
         \includegraphics[width=\textwidth]{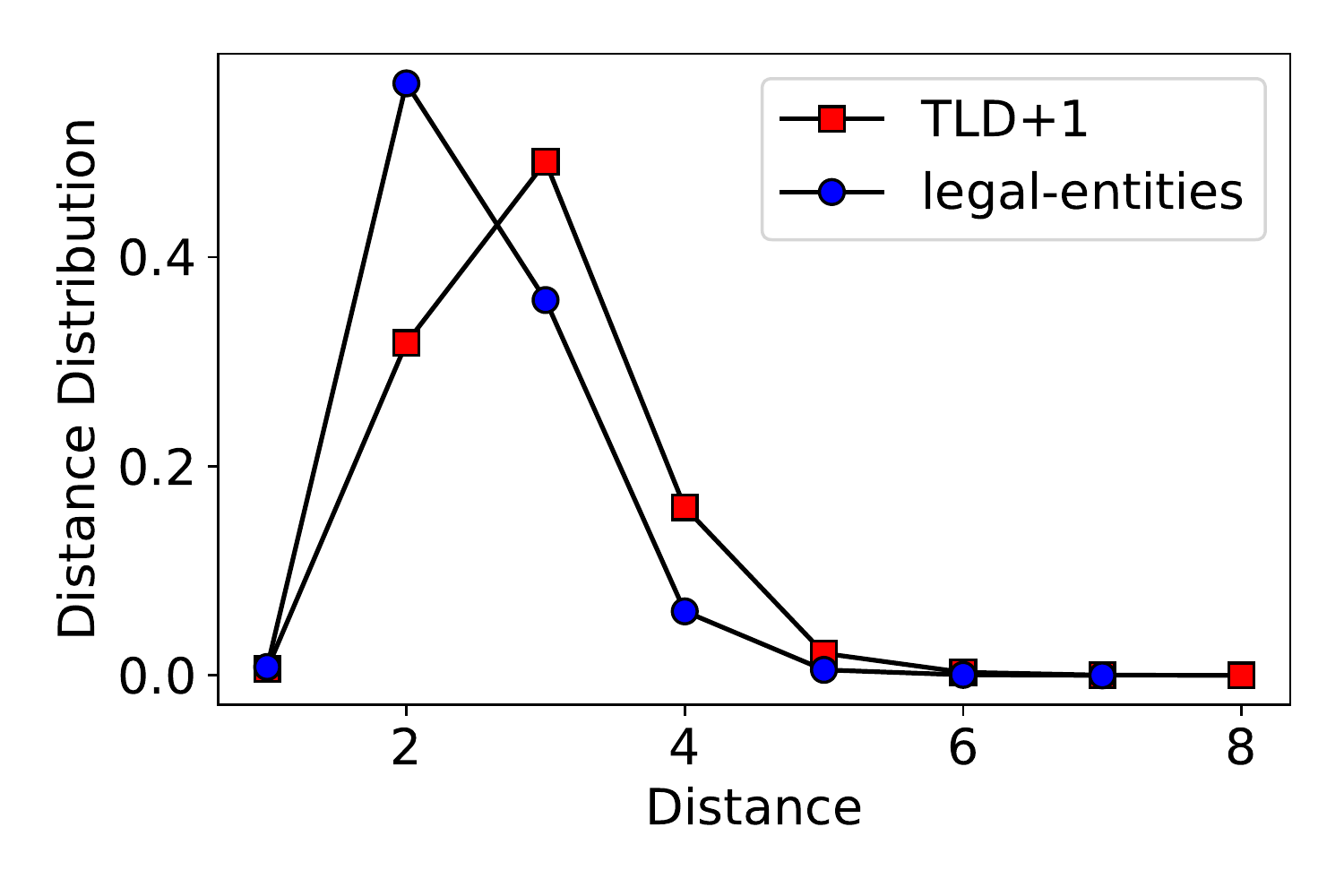}
         \vspace{-7mm}
         \caption{Distance distribution $d(l)$}
         \label{subfig:distance-distribution}
     \end{subfigure}
     \begin{subfigure}[b]{0.31\textwidth}
         \centering
         \includegraphics[width=\textwidth]{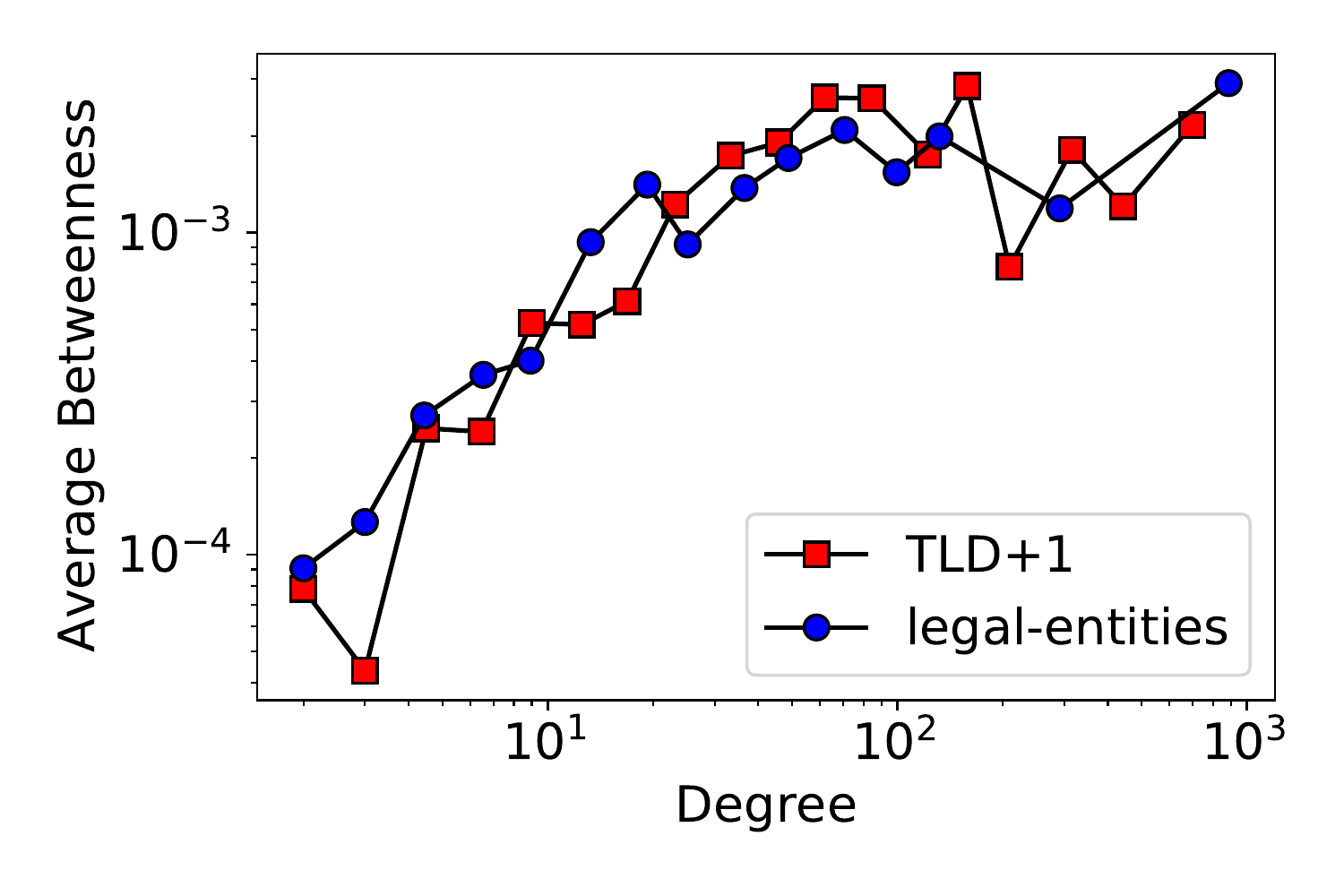}
         \vspace{-7mm}
         \caption{Average betweenness $\bar{B}(k)$}
         \label{subfig:average-betweenness}
     \end{subfigure}
     \vspace{-3mm}
    \caption{Properties of the TLD+1 and legal-entity networks.}
    \label{fig:structural-properties}
\end{figure*}

In Fig.~\ref{fig:structural-properties} we also compute the following properties, as in~\cite{Papadopoulos2012}: (a) the degree distribution $P(k)$; (b) the average clustering $\bar{c}(k)$ of $k$-degree nodes; (c) the average neighbor degree $\bar{k}_{\text{nn}}(k)$ of $k$-degree nodes; (d) the distance distribution $d(l)$, i.e., the distribution of hop lengths $l$ between nodes in the network; and (e) the average node betweenness $\bar{B}(k)$ of $k$-degree nodes, which is the average number of shortest paths passing through a $k$-degree node, normalized by the maximum possible number of such paths. Properties (a-c) are local statistics reflecting properties of individual nodes and their one-hop neighborhoods, as opposed to global properties (d,~e). From the figure we see that these properties are remarkably similar between the TLD+1 and legal-entity networks, despite the latter being an aggregation of the former. The degree distribution in both cases can be approximated by a power-law, $P(k) \propto k^{-\gamma}$, with exponent $\gamma \approx 2.3$.

\begin{table}[t]
\caption{Overview of the \textbf{TLD+1} and legal-entity networks}
\vspace{-3mm}
\label{tab:networkGraphsStats}
\resizebox{\columnwidth}{!}{%
\begin{tabular}{r|c|c|}
\cline{2-3}
& \textbf{TLD+1} & \textbf{legal-entities} \\ \hline
\multicolumn{1}{|r|}{\textbf{Number of nodes $N$}} & 1847 & 1215 \\ \hline
\multicolumn{1}{|r|}{\textbf{Avg. clustering coefficient $\bar{c}$}} & 0.52 & 0.59 \\ \hline
\multicolumn{1}{|r|}{\textbf{Avg. node degree $\bar{k}$}} & 11.54 & 7.74 \\ \hline
\multicolumn{1}{|r|}{\textbf{Max. node degree $k_\textnormal{max}$}} & 728 & 886 \\ \hline
\multicolumn{1}{|r|}{\textbf{Power-law exponent $\gamma$}} & 2.29 & 2.28\\ \hline
\multicolumn{1}{|r|}{\textbf{Avg. distance $\bar{d}$}} & 2.88 & 2.49 \\ \hline
\multicolumn{1}{|r|}{\textbf{Max. distance $d_\textnormal{max}$}} & 8 & 7 \\ \hline
\end{tabular}
}
\end{table}

To see how the number of users contributing data affects the inferred TLD+1 network, we perform the following process. We sort the users according to the total number of TLD+1 interactions they collect, from highest to lowest. Then, starting from the first user we sum the number of new TLD+1 nodes introduced to the network per additional user, and normalize this sum with the total number of TLD+1 nodes in the network. We perform the same process for the number of new TLD+1 links instead of nodes. The results are shown in Fig.~\ref{fig:NodesPerUser}. We see that with only the first 100 users we discover more than 95\% of the total TLD+1 nodes. To discover 95\% of the total TLD+1 links we need to consider the first 200 users. Taken altogether, these results show that with less than half of the considered users we can discover a vast proportion of our TLD+1 network, while considering additional users does not have a significant effect on the structure of the network.

We note that we also constructed corresponding TLD+1 networks for five different continents (Europe, South America, Asia, Africa and North America) and found that they have qualitatively similar topological characteristics as that of the global TPD network and exhibit high percentages of common nodes and links (see Appendix~\ref{appendix_continent}).

It has been found during the last decade that hyperbolic geometry underlies the topology of many real-world complex networks, naturally explaining their strong clustering and heterogeneous degree distributions~\cite{Krioukov2010,Boguna2010,Papadopoulos2012}. Node coordinates in the underlying or ``hidden'' space can be inferred from the observed network using machine learning and statistical inference methods~\cite{Boguna2010, Papadopoulos_2015, Garc_a_P_rez_2019}. The node coordinates abstract how popular and similar the nodes are~\cite{Papadopoulos2012}, while hyperbolically closer nodes have higher chances of being connected in the observed network~\cite{Krioukov2010}. In the next section we map the legal-entity network into its underlying hyperbolic space and analyze the resulting embedding.

\section{Hyperbolic Embedding of the legal-entity network} \label{sec:hyperbolic}

\subsection{Methodology}

We map the legal-entity network using the recently developed Mercator method~\cite{Garc_a_P_rez_2019}. Mercator takes as input the network adjacency matrix $\alpha_{ij}$---$\alpha_{ij}=\alpha_{ji}=1$ if there is a link between nodes $i$ and $j$, and $\alpha_{ij}=\alpha_{ji}=0$ otherwise, and computes radial and angular coordinates $r_i, \theta_i$, for all nodes $i=1, \ldots, N$. The radial coordinate $r_i$ abstracts node's $i$ popularity. The smaller the $r_i$ the larger is the expected degree of node $i$. On the other hand, angular coordinates abstract node similarities. The smaller the angular distance $\Delta\theta_{ij}=\pi-|\pi-|\theta_i-\theta_j||$ is between two nodes $i$ and $j$ the more similar the two nodes are. The hyperbolic distance $x_{ij}$ between two nodes $i$ and $j$ is a single-metric representation of the nodes' popularity and similarity attributes, and is given by the law of hyperbolic cosines~\cite{Krioukov2010}:
\begin{equation}
x_{ij} =\mathrm{arccosh}\left(\cosh{r_i}\cosh{r_j}-\sinh{r_i} \sinh {r_j} \cos{\Delta\theta_{ij}}\right).
\end{equation}
In a nutshell, Mercator infers the nodes' radial and angular coordinates by maximizing the likelihood:
\begin{equation}
\label{eq:likelihood}
\mathcal L=\prod_{1 \leq j < i \leq N} p(x_{ij})^{\alpha_{ij}}\left[1-p(x_{ij})\right]^{1-\alpha_{ij}},
\end{equation}
where the product goes over all node pairs $i, j$ in the network, while $p(x_{ij})$ is the Fermi-Dirac connection probability~\cite{Krioukov2010} between nodes $i$ and $j$:
\begin{equation}
\label{eq:pxij}
p(x_{ij})=\frac{1}{1+e^{\frac{1}{2T}(x_{ij}-R)}}.
\end{equation}
In the last relation, $R \propto \ln{N}$ is the radius of the hyperbolic disk where all nodes reside, while $T \in (0,1)$ is a parameter called ``network temperature'' that is also inferred by Mercator and is related to the amount of clustering in the network. We note that the connection probability $p(x_{ij})$ decreases with the hyperbolic distance $x_{ij}$ between nodes $i, j$. Thus, maximization of the likelihood in Eq.~(\ref{eq:likelihood}) corresponds to attracting connected pairs of nodes closer in the hyperbolic space, while repelling disconnected ones. This interplay between attraction and repulsion effectively maps the nodes belonging to densely connected groups closer in the hyperbolic space. See~\cite{Boguna2010, Garc_a_P_rez_2019} for further details.

The code implementing Mercator is available at~\cite{Mercator_repo}. We use the code as is without any modifications.

\begin{figure*}[ht]
\centering
\includegraphics[width=\textwidth]{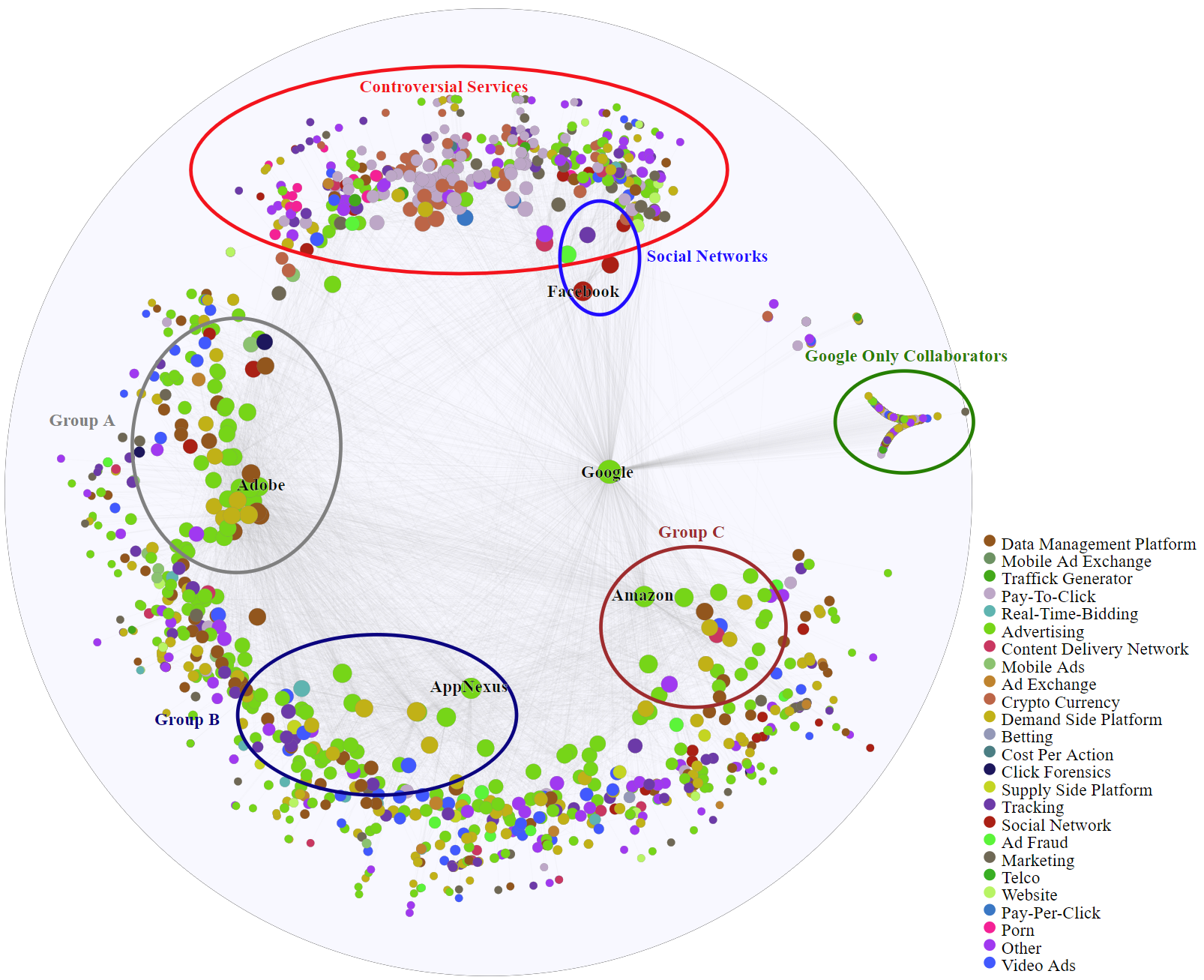}
\vspace{-6mm}
\caption{Hyperbolic map of the legal-entity network.}
\label{fig:hypermap}
\end{figure*}

\subsection{Hyperbolic map of the legal-entity network} \label{sec:visualization}

Figure~\ref{fig:hypermap} visualizes the hyperbolic embedding of the legal-entity network. Node colors indicate the type of the online activities of the corresponding legal entities as listed in the figure's legend, while the size of the nodes is proportional to the logarithm of their degree.

The largest hub (node with largest degree) is Google. It has a dominant role, sitting almost in the center of the map and interconnecting different communities (groups of nodes close along the angular similarity direction). On the right side of the map we see a small group of nodes (Google Only Collaborators) that are only connected directly with Google. This group includes organizations that provide services related to online advertising, such as Video Advertising, Demand Side Platforms, Data Management Platforms, Mobile Ads, etc.
On the middle left and lower half of the map we can identify three groups (Groups A, B, C) whose nodes provide services related to online advertising, like the ones in the Google Only Collaborators group. For readability, we report on the map only the name of the legal entity with the largest degree in each group, which is Adobe, AppNexus and Amazon for groups A, B and C, respectively. Higher-degree nodes in these three groups are ad exchanges, ad networks and nodes dedicated to advertising, while smaller-degree nodes are Demand Side Platforms, Data Management Platforms, and Trackers. These observations regarding these three groups are in line with prior work focusing on the tracking and advertising TPD ecosystem~\cite{Bashir2018, Gomer2013, Solomos2019}, and indicate that the hyperbolic map is meaningful.

On the top of the map we observe two groups that partially overlap: the Controversial Services group and the Social Networks group. The Controversial Services group includes legal entities that provide controversial services, such as Pay-To-Click (PTC), Crypto Currency, Betting, and Porn. 

We have manually visited and inspected the websites of domains under each legal entity of the Controversial Services group in order to better understand their activities and interactions. Overall we manually inspect 271 domains. The main activities of PTC nodes is to act as middlemen between advertisers and consumers. PTCs display ads from advertisers and earn money whenever an ad is clicked on by a viewer. In return, a fraction of the fee goes to the viewer. By contrast, in the Pay-Per-Click (PPC) model only the publisher of the ad will be paid for the ad clicks. The Crypto Currency nodes include activities related to cryptoadvertising, that is, advertisement based on cryptocurrency payments, which is usually preferred in PTC. The Porn related nodes provide adult content videos and advertising services that can be also combined with cryptocurrency payment in order to maintain anonymity of the involved actors.

The Social Networks group includes the three main social networks, Facebook, Twitter and LinkedIn. To understand its partial overlap with the Controversial Services group we performed a further manual investigation of the websites in the Controversial Services and found the following: (1) most PTCs are utilizing cryptocurrencies to pay their ad viewers; (2) PTCs do not report any physical address of their organization's headquarter nor they provide any support through their websites; and (3) PTCs utilize social network campaigns and have a strong social network presence in order to recruit viewers and provide support to them. We therefore find evidence that controversial advertising sites heavily depend on and utilize social networks, justifying the partial overlap we see between the Controversial Services and Social Networks groups in the hyperbolic map (Fig.~\ref{fig:hypermap}).

To our best knowledge, the place and integration of controversial advertising services within the TPD ecosystem, as well as their close interactions with social networks, revealed by the hyperbolic embedding of the ecosystem, have not been revealed nor analyzed in prior work. Therefore, the hyperbolic map of the ecosystem is not only meaningful, but can also lead to new insights and motivate further investigations. We make our visualization tool and data used to construct Fig.~\ref{fig:hypermap} publicly available~\cite{TPD_repo}. Our visualization tool allows interaction with the hyperbolic map, e.g., zoom in on regions of interest, selection of nodes to display more information, etc. (see Appendix~\ref{app:visual_map} for further details).

\section{TLD+1 Grouping and Legal Entities Co-hosting} \label{sec:applications}

\begin{figure*}[ht]
     \centering
     \begin{subfigure}[b]{0.49\textwidth}
         \centering
         \includegraphics[width=\textwidth]{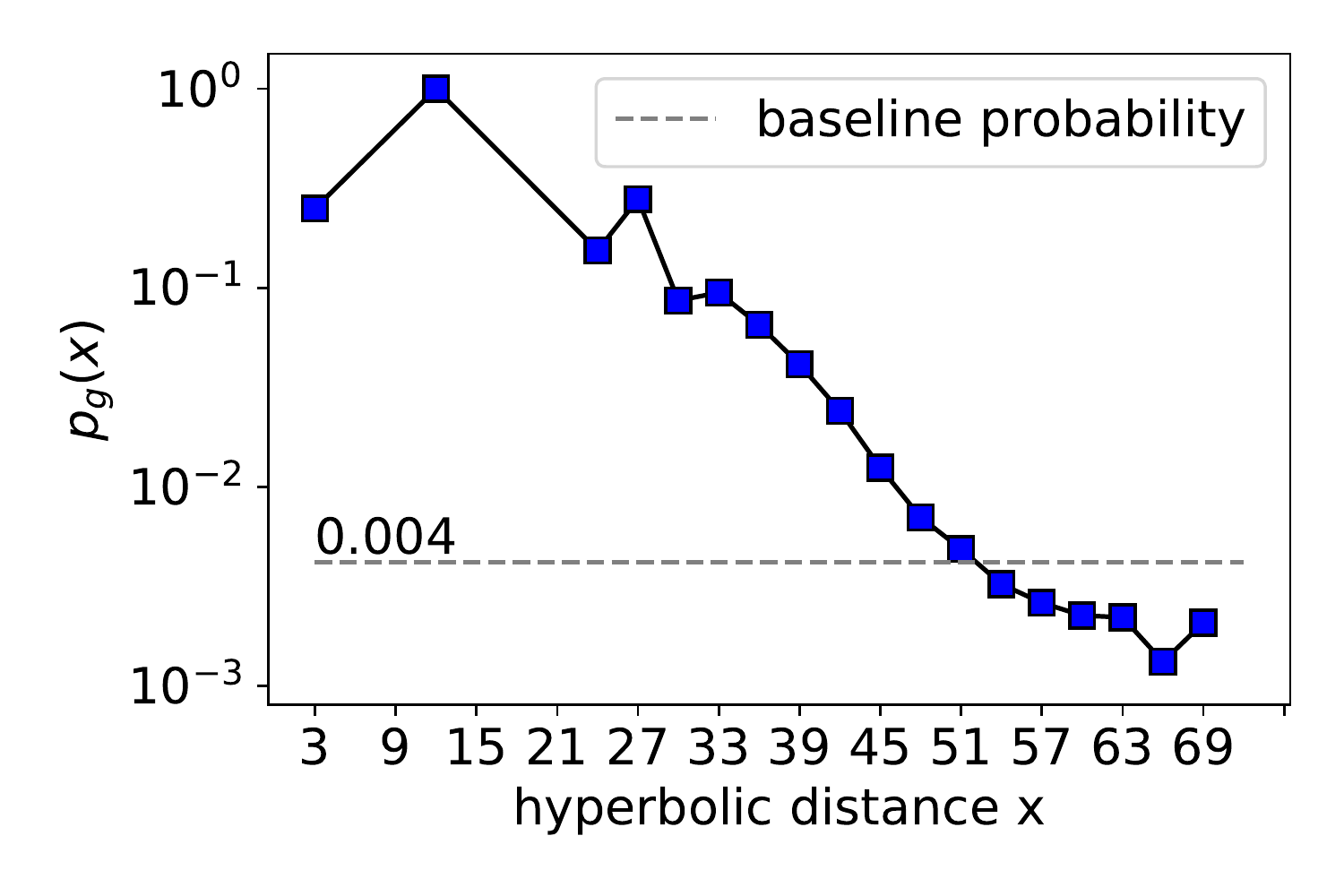}
         \vspace{-9mm}
         \caption{}
         \label{fig:Merge-hyp-Organizations}
     \end{subfigure}
     \hfill
     \begin{subfigure}[b]{0.49\textwidth}
         \centering
         \includegraphics[width=\textwidth]{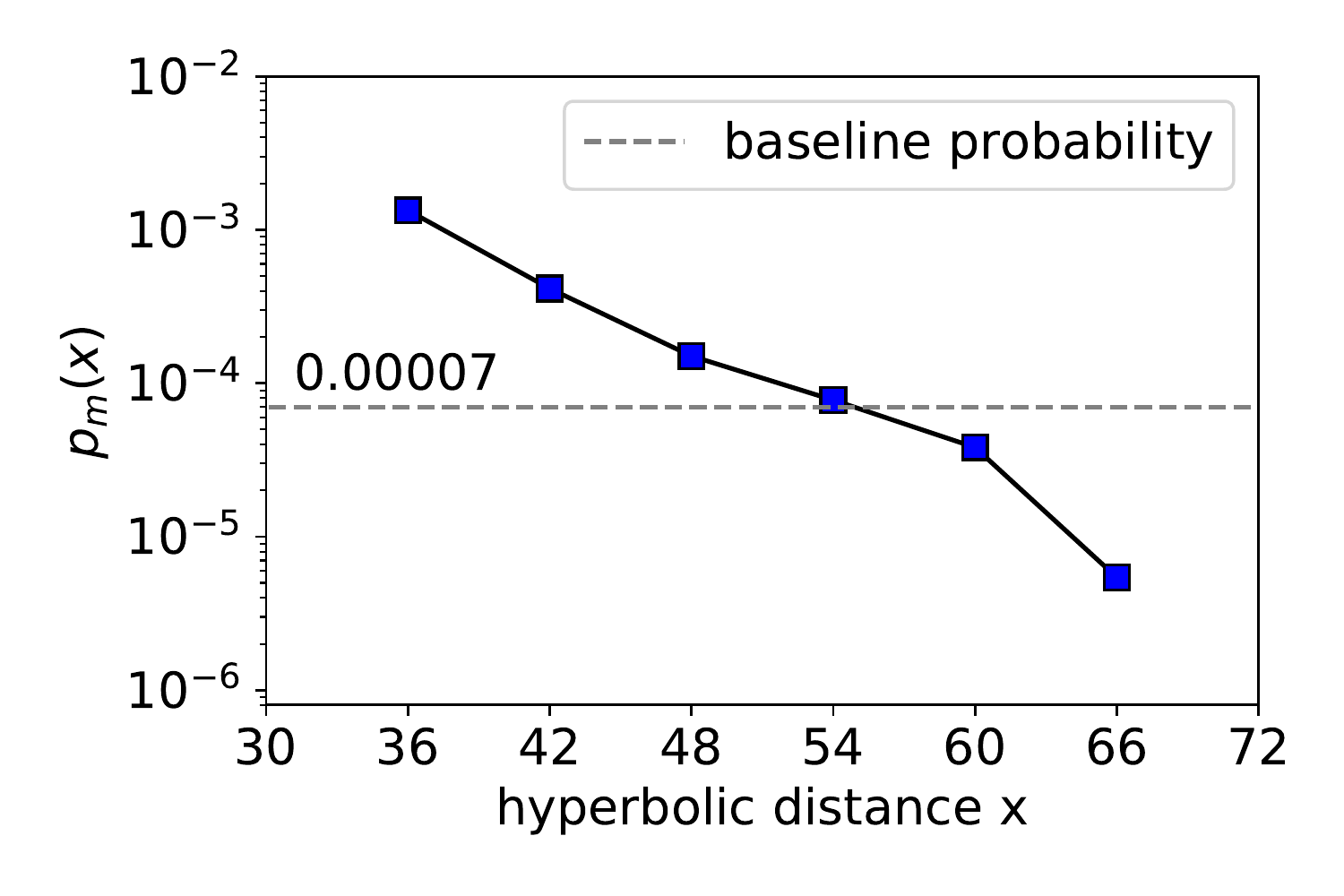}
         \vspace{-9mm}
         \caption{}
         \label{fig:future-hyp-organizations}
    \end{subfigure}
    \vspace{-2mm}
    \caption{(a) Grouping probability $p_{g}(x)$ as a function of the hyperbolic distance $x$ between TLD+1 nodes. (b) Future merging probability $p_m(x)$ as a function of the hyperbolic distance $x$ between TLD+1 nodes. The horizontal dashed line in each case indicates the corresponding baseline probability.}
        \label{fig:Grouping-hyp-Organizations}
\end{figure*}

\subsection{TLD+1 Grouping} 
\label{sec:merging}


\begin{figure*} 
     \centering
     \begin{subfigure}[b]{0.49\textwidth}
         \centering
         \includegraphics[width=\textwidth]{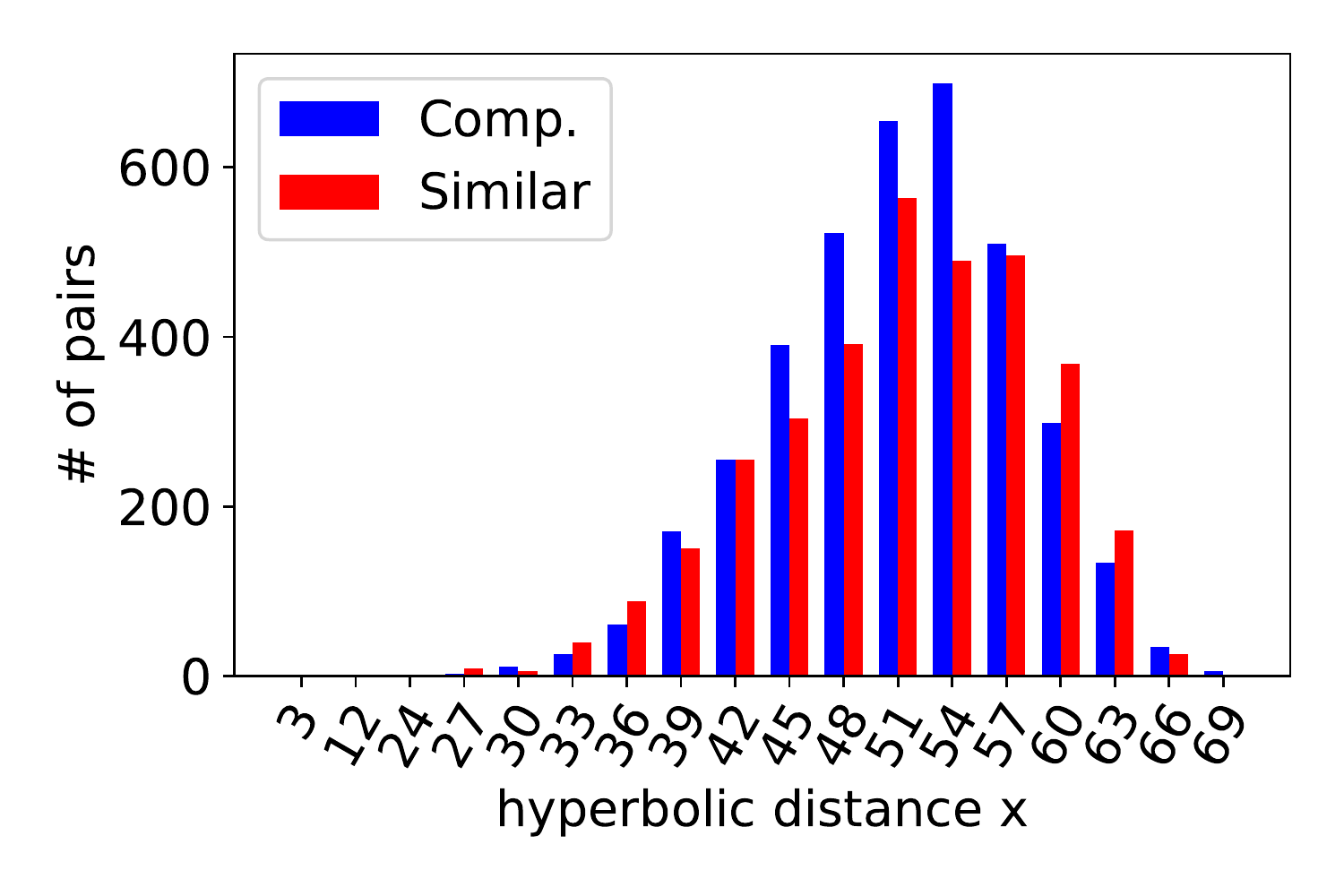}
         \vspace{-9mm}
         \caption{}
         \label{subfig:comp-vs-sim-all}
     \end{subfigure}
     \hfill
     \begin{subfigure}[b]{0.49\textwidth}
         \centering
         \includegraphics[width=\textwidth]{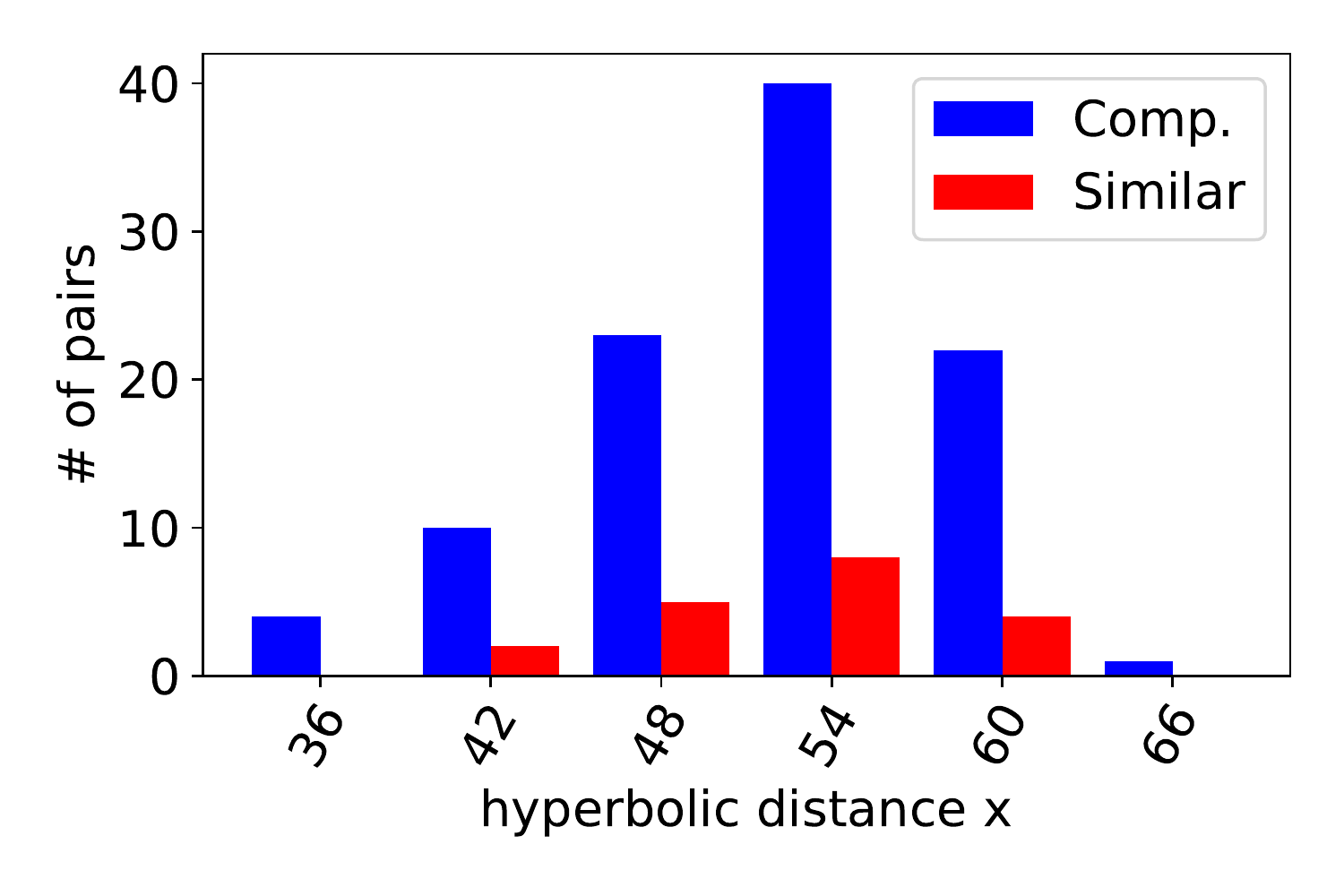}
         \vspace{-9mm}
         \caption{}
         \label{subfig:comp-vs-sim-future}
     \end{subfigure}
     \vspace{-2mm}
    \caption{(a) Complementarity (Comp.) Vs. Similarity of TLD+1 pairs that belong to the same entity. (b) Same as in (a) but for TLD+1 pairs that merge under the same entity in the future (see text).}
    \label{fig:comp-vs-sim}
\end{figure*}

Our analysis here is motivated by recent legislations, such as the GDPR~\cite{gdpr}, which put pressure on TPDs forcing them to follow new standards and provide more transparency about their online activities and data handling, which can be translated into higher operational costs for the TPDs. Furthermore, TPDs are in a constant hanger for more data about users' online activities. The combination of these two factors can force TPDs to group or merge their existing networks in the form of company acquisitions in an attempt to mitigate the above legal issues and at the same time increase their user base and tracking capabilities~\cite{postGDPR}. Given these considerations, we investigate if the hyperbolic distances among TPDs give information about the likelihood of their current grouping or future merging. Specifically, we hypothesize that hyperbolically closer TPDs are more likely to belong to the same legal entity, or merge under the same legal entity in the future. To verify these hypotheses we map the TLD+1 network into its underlying hyperbolic space using Mercator~\cite{Garc_a_P_rez_2019} and compute empirical \emph{grouping} and \emph{future merging} probabilities as a function of hyperbolic distance, as we describe below.

\subsubsection{Current Grouping}
\label{sec:grouping}
The empirical grouping probability is the probability that a pair of TLD+1 nodes belongs to the same legal entity given the pair's hyperbolic distance $x$. This probability is denoted by $p_g(x)$, and computed as follows. First, we compute the hyperbolic distances among all node pairs. We then bin the range of hyperbolic distances from zero to the maximum distance into small bins. For each bin we find all the pairs located at the hyperbolic distances falling within the bin. The percentage of pairs in this set of pairs that belong to the same legal entity is the value of the corresponding grouping probability at the bin. (The correspondence of TLD+1 nodes to legal entities is given by the legal-entity list constructed in Sect. \ref{subsec:data_processing}.) We note that the TLD+1 network consists of $N=1847$ nodes, and thus $N(N-1)/2=1704781$ node pairs. From these pairs, $7140$, i.e., approximately $0.42\%$ belong to the same entity. This percentage is the probability that a pair of TLD+1 nodes belongs to the same entity if we do not know its hyperbolic distance. We call this probability \emph{baseline probability}.

Figure~\ref{fig:Merge-hyp-Organizations} shows the results. We see that $p_g(x)$ tends to decrease with the hyperbolic distance $x$ between TLD+1 nodes, verifying our hypothesis that hyperbolically closer TPDs are more likely to belong to the same entity. Overall, we observe that $p_g(x)$ is between two to three orders of magnitude higher for pairs at smaller hyperbolic distances compared to pairs at larger distances. Fig.~\ref{fig:Merge-hyp-Organizations} also juxtaposes $p_g(x)$ against the baseline grouping probability, illustrating the discriminatory power of the embedding.

\subsubsection{Future Merging}
\label{sec:future_predictions}
Here, we compute the empirical future merging probability $p_{m}(x)$ as a function of the hyperbolic distance $x$ between TLD+1 nodes. This probability is computed for the period April 2020 to November 2021, which is after the period in which we constructed the TLD+1 network. To this end, we first bin the TLD+1 pairs according to their hyperbolic distance as in Sect.~\ref{sec:grouping}. Then, for each bin we find the percentage of pairs in the bin that merge under the same legal entity in the aforementioned period. (The merging of TLD+1 nodes to entities in the above period is given by the future-merging list constructed in Sect.~\ref{subsec:data_processing}.) We note that we have observed a total of $119$ TLD+1 pairs merging under the same entity. This corresponds to approximately $0.007\%$ of all TLD+1 pairs. This percentage represents the baseline merging probability. We also note that we ignore bins where the estimated merging probability is zero, as this estimation is not expected to be accurate given the low number of observed mergings. These are mainly bins at low distances with relative few node pairs (less than $0.2\%$ of all pairs).

Figure~\ref{fig:future-hyp-organizations} shows the results. Similar to Fig.~\ref{fig:Merge-hyp-Organizations}, we observe that $p_m(x)$ decreases with the hyperbolic distance $x$ between TLD+1 nodes, verifying our hypothesis that hyperbolically closer TPDs are more likely to merge under the same legal entity in the future.
Again, $p_m(x)$ is between two to three orders of magnitude higher for pairs at smaller hyperbolic distances compared to pairs at larger distances. Fig.~\ref{fig:future-hyp-organizations} also shows the corresponding baseline probability, illustrating the discriminatory power of the embedding.


\subsubsection{Complementarity Vs. Similarity} 
\label{subsub:complimentarity}

For each TLD+1 node in our data, we also recorded its online activities (Sect.~\ref{subsec:data_processing}). Having this information available, we can directly check whether TLD+1 pairs that belong to the same entity, or merge in the future under the same entity, are characterized by similar or complementary services. Specifically, we call here ``similar'' a pair of nodes that provides similar services, e.g., web tracking. Further, we call ``complementary'' a pair of nodes that provides complementary services, e.g., one node provides services related to web tracking and the other node provides services related to web advertising.

Figure~\ref{subfig:comp-vs-sim-all} shows the distribution of the number of similar and complementary pairs of TLD+1 nodes as a function of their hyperbolic distance. The figure considers all pairs that belong to the same entity at the end of our measurement period (7140 pairs in total). We see that irrespective of their hyperbolic distance, TLD+1 pairs belonging to the same entity have comparable chances of being similar or complementary. On the other hand, Fig.~\ref{subfig:comp-vs-sim-future} considers the TLD+1 pairs that merge under the same entity after our measurement period (that is, in April 2020 to November 2021). We see from Fig.~\ref{subfig:comp-vs-sim-future} that these future mergings are predominated by complementary TLD+1 pairs. These results suggest that complementarity instead of similarity is the dominant force driving TPD mergings (Fig.~\ref{subfig:comp-vs-sim-future}), however, after merging, TPDs tend to become more similar (Fig.~\ref{subfig:comp-vs-sim-all}).


\subsection{Legal entities co-hosting} \label{sec:co-hosting}
Finally, we consider legal entities co-hosting. Legal entities may be co-hosted in an attempt to facilitate more efficient exchange of information between them, such as fast exchange of data related to users' online activities~\cite{srinivasan2020google, Doh-Shin2021}. Such practice is common in the advertising ecosystem due to the time constrains imposed by the Real-Time Bidding protocol~\cite{openRTB, srinivasan2020google, Doh-Shin2021}. Therefore, we consider entity co-hosting an indicator for possible collaboration among the entities, and hypothesize that hyperbolically closer entities have higher chances of being co-hosted.

\begin{figure}[ht] 
\centering
\includegraphics[width=\columnwidth]{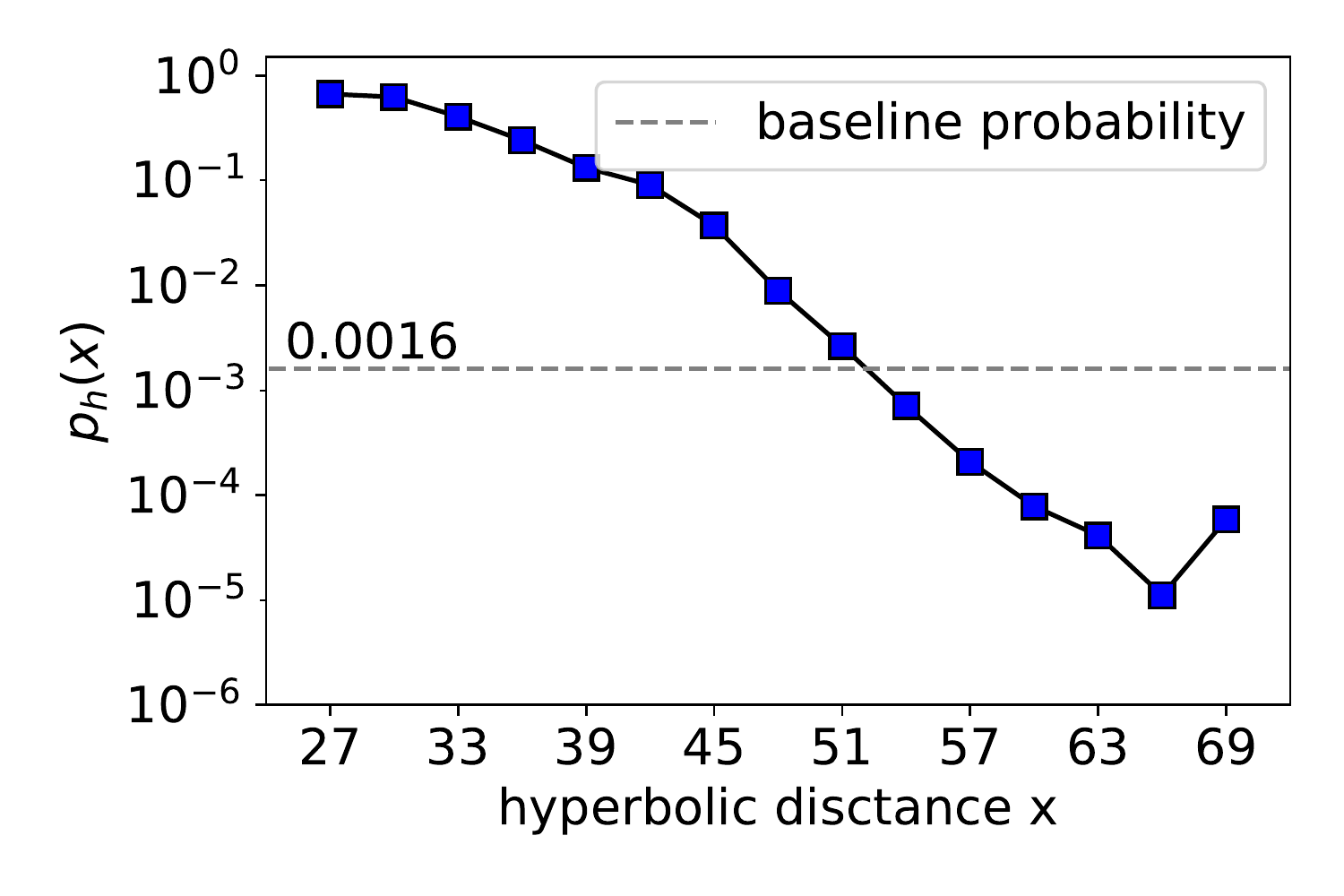}
\vspace{-2mm}
\caption{Co-hosting probability $p_h(x)$ as a function of the hyperbolic distance $x$ between legal entities.}
\label{fig:co-hosted-hyp-organizations}
\end{figure}

To verify this hypothesis, we consider the hyperbolic embedding of the legal-entity network from Sect.~\ref{sec:hyperbolic}, and compute the empirical \emph{co-hosting probability} $p_h(x)$ as a function of the hyperbolic distance $x$ between the entities, following a similar binning procedure as in Sect.~\ref{sec:merging}. (The information about which entities are co-hosted is provided by the co-hosting list constructed in Sect. \ref{subsec:data_processing}.)
The results are shown in Fig.~\ref{fig:co-hosted-hyp-organizations} and are qualitatively similar to the results in Fig.~\ref{fig:Merge-hyp-Organizations}. Specifically, we see that $p_h(x)$ decreases with the hyperbolic distance $x$ between entities, verifying that hyperbolically closer entities are more likely to be co-hosted. We note that the legal-entity network consists of $N=1215$ nodes, and thus $N(N-1)/2=737505$ node pairs. From these pairs, $1181$, i.e., $0.16\%$ are co-hosted. This percentage represents the baseline co-hosting probability, i.e, the probability that two entities are co-hosted if we do not know the hyperbolic distance between them. Fig.~\ref{fig:co-hosted-hyp-organizations} also depicts the baseline co-hosting probability, illustrating again the discriminatory power of the embedding.

Therefore, hyperbolic embedding of the TPD ecosystem provides us with inferential power at different aggregation levels of the ecosystem. Legal entity co-hosting and TLD+1 grouping or future merging may imply collaboration between the involved entities and TPDs~\cite{openRTB, srinivasan2020google, Doh-Shin2021}. Thus, the hyperbolic map of the ecosystem could be a useful tool in the arsenal of regulators and data protection authorities for investigations related to TPDs that may be collaborating, and for which no ground-truth data is available that indicates their collaboration.


\section{Interaction paths and navigability} 
\label{sec:communication}

In this section we turn our attention to interaction paths and the navigability of the TPD ecosystem. As explained in Sect.~\ref{subsec:data_collection}, during the rendering of a website the first-party domain may invoke different TPDs, which in turn invoke other TPDs, and so on, thus creating chains of interactions between TPDs. We call each interaction chain that starts from one TPD (``source'') and ends at another TPD (``destination'') an \emph{interaction path}. We note that different interaction paths between the same source-destination TPDs may be observed by monitoring different websites.

\begin{figure}[!b] 
\centering
\includegraphics[width=0.95\columnwidth]{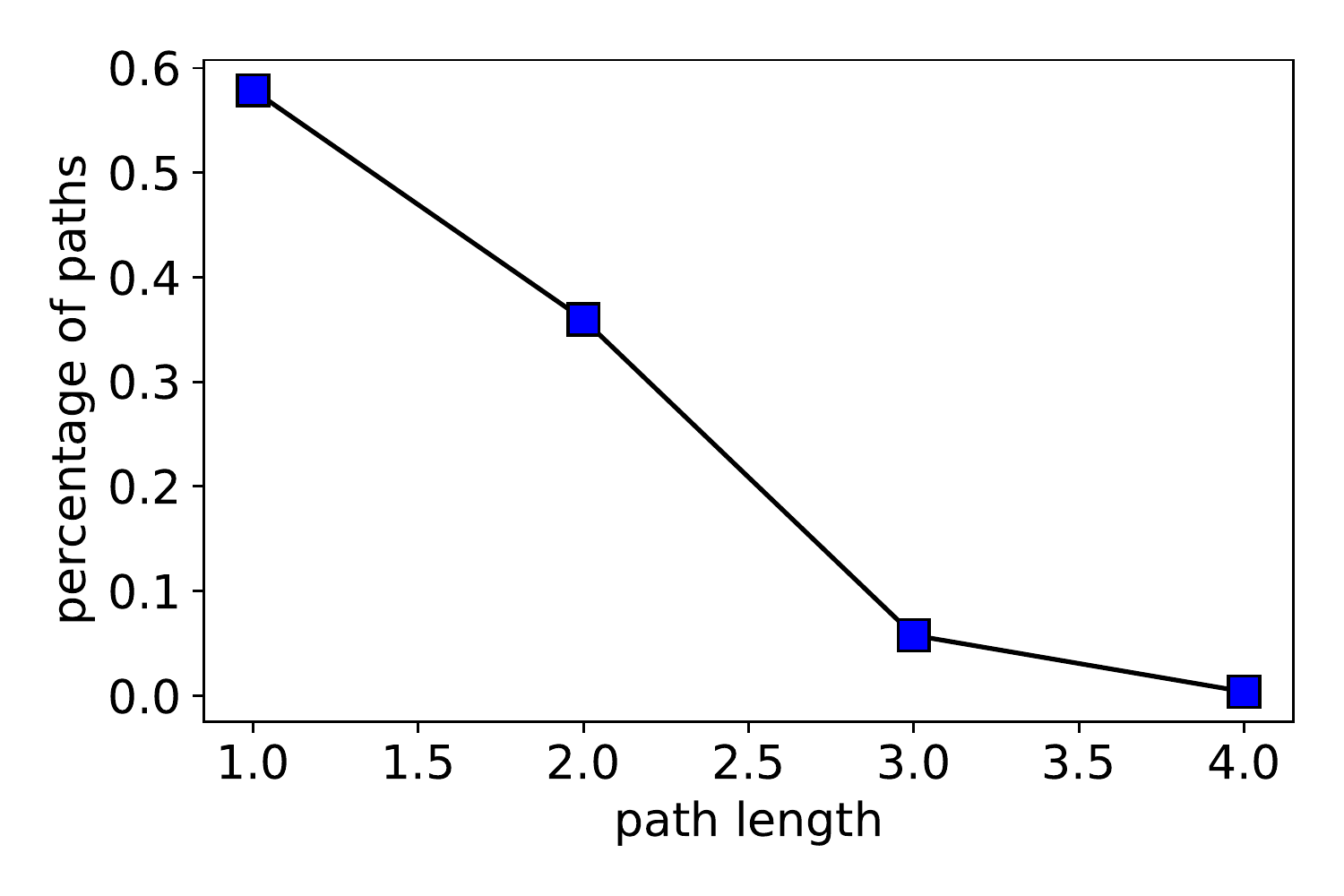}
\vspace{-7mm}
\caption{Distribution of interaction path lengths between legal entities.
}
\label{fig:actual_pdf}
\end{figure}

Knowing the legal entity to which each TPD belongs to (given by the legal-entity list of Sect.~\ref{subsec:data_processing}) we can find the interaction paths between the corresponding legal entities. Overall, we identify 10747 such paths that correspond to 7354 different source-destination entity pairs. Figure~\ref{fig:actual_pdf} shows the distribution of the hop-lengths of these paths. These interaction paths traverse only around 1.5 hops on average, while the maximum number of hops is four. Overall, we find that around $97$\% of the interaction paths follow shortest paths in the legal-entity network.

\textbf{Navigability.} The navigability of a network~\cite{Boguna2007} can be used as an alternative indirect metric of embedding quality~\cite{Papadopoulos_2015, Papadopoulos_2015_2}. Navigability of an embedding is also of independent interest for some applications, such as Internet routing~\cite{Boguna2010}. A network embedded in a geometric space is said \emph{navigable} if \emph{greedy routing (GR)} is efficient according to the metrics considered below. In GR, a node's address is its coordinates in the space, and each node knows only the addresses of its neighbors, and the destination node address of a ``packet''. Upon receipt of such a packet, the GR node, if it is not a destination, forwards the packet to its neighbor closest to the destination in the geometric space, and drops the packet if a local minimum loop is detected, i.e., if this neighbor is the same as the previous node visited by the packet.

We evaluate the efficiency of GR in the legal-entity and TLD+1 networks using the nodes' inferred hyperbolic coordinates in each case. We consider the following GR efficiency metrics: (i) the percentage of successful paths, $p_s$, which is the proportion of paths that do not get looped and reach their destinations; and (ii) the average and maximum stretch of successful paths, denoted by $\bar{s}$ and $s_\text{max}$ respectively. The stretch is defined as the ratio between the hop-lengths of greedy paths and the corresponding shortest paths in the network. We run the GR process between all possible source-destination pairs in each network. We find that in the legal-entity network $p_s=0.91$, $\bar{s}=1.02$ and $s_\text{max}=2.0$. Thus, GR is highly successful while greedy paths follow almost shortest paths in the network. These results show that the legal-entity network is highly navigable, and provide a further validation of the quality of the embedding in Sect.~\ref{sec:hyperbolic}. The TLD+1 network is also well-navigable but not as navigable as the legal-entity network. In particular, in the TLD+1 network we find $p_s=0.75$, $\bar{s}=1.08$ and $s_\text{max}=2.67$.

\section{Related Work}
\label{sec:related}

To our best knowledge, no prior work has captured and analyzed the topology of the TPD ecosystem at a global scale. This may not be surprising given the associated technical challenges in tracking TPD interactions at a large-scale, described in Sect.~\ref{subsec:Challenges}. As explained (Sect.~\ref{subsec:Challenges}), to overcome these challenges we directly tracked TPD interactions from real users' browsers from all over the world. In this context, the most related prior studies are the ones by Gomer et al.~\cite{Gomer2013} and Bashir et al.~\cite{Bashir2018}. However, these studies focus only on the tracking and advertising subsystem of the global TPD ecosystem. They both utilize data collected via web crawling, and follow different approaches for constructing the tracking and advertising network. They observe that at the TLD+1 level this network exhibits the small-world property, has high clustering and high-degree hubs, and is disassortative (high-degree nodes tend to connect to low degree nodes). We have observed similar properties for the network of the global TPD ecosystem in Sec.~\ref{sec:graph}. Furthermore, we have shown for the first time that the degree distribution in the global TPD network can be well-approximated by a power law.  

On the other hand, there has been plenty of work in the area of network geometry during the last decade, partly motivated by the seminal paper of Krioukov et al.~\cite{Krioukov2010}. Krioukov et al. showed that hyperbolic geometry is the natural geometry underlying the topology of real-world networks, which are characterized by heterogeneous degree distributions and strong clustering. In another seminal work, Bogu{\~{n}}{\'{a}} et al.~\cite{Boguna2010} developed the first methodology to map real complex networks into their intrinsic hyperbolic spaces. The authors showed that the maps are meaningful and can facilitate efficient greedy routing and geometric community detection in the Internet. Papadopoulos et al.~\cite{Papadopoulos2012} elucidated that the hyperbolic node coordinates in network embeddings abstract the popularities and similarities of nodes. Furthermore, it has been shown that hyperbolic embeddings can facilitate the prediction of missing and future links in networks~\cite{Papadopoulos_2015, Papadopoulos_2015_2}. Finally, several approaches for inferring hyperbolic network coordinates (with different tradeoffs between computational complexity and accuracy) have been developed~\cite{Papadopoulos_2015, Papadopoulos_2015_2, blasius16, carlo2017, Garc_a_P_rez_2019}. For a recent review on network geometry see Ref.~\cite{Boguna2021}. No prior work has investigated whether the global TPD ecosystem has an underlying hyperbolic geometry and whether its large-scale network admits a meaningful hyperbolic embedding.

\section{Conclusion} 
\label{sec:conclusions}

We have analyzed the large-scale network of the TPD ecosystem. To track this network we monitored the interactions between TPDs in real users' browsers from all over the world. Our inferred network may not be complete but it is the largest one studied to date. All known TPDs reported in prior work constitute a subset of the TPDs we tracked here. We release our data and related code to the public~\cite{TPD_repo}.

We have considered two levels of aggregation of the TPD ecosystem (TLD+1 and legal-entity levels) and found that the resulting networks possess similar structural properties, commonly found in complex networks. Given these properties, we have shown that the TPD ecosystem admits meaningful hyperbolic embeddings, showcasing the large-scale organization of the ecosystem and revealing interactions between TPDs. In particular, we have observed strong ties between controversial services and social networks, which warrant further investigation as part of future work (e.g., what is the exact nature of the interactions between the two groups, what kind of information is exchanged, are interactions bidirectional, etc.). Furthermore, we have shown that the hyperbolic embedding of the ecosystem possesses inferential and predictive power. In particular, we have shown that it can provide information about TPDs current groupings and future mergings and legal entities' co-hostings. Thus, it could be a useful tool for conducting investigations related to TPDs that may be collaborating and for which no ground-truth data is available that indicates their collaboration. We also found evidence that complementarity instead of similarity is the dominant force driving future TPD mergings. Further, we have tracked actual interaction paths between legal entities and found that these paths almost always follow shortest paths in the legal-entity network. In addition, we found that the legal-entity and TLD+1 networks are well-navigable in their hyperbolic embeddings.

It is also possible that the embedding of the TPD ecosystem could be applicable in other contexts, such as inferring first-party cookies sharing (stolen cookies)~\cite{Quan2021} and cloud service sharing~\cite{Iordanou_2018} between TPDs. We also note that mergings that can lead to monopolies~\cite{Beatriz2021} will be prohibited according to a recent report~\cite{us_monopolies} by the U.S House Judiciary Committee. Therefore, methodologies for identifying probable TPD mergings, like the one we considered here, could be relevant to such contexts. Further, we note that while we focused here on the global TPD ecosystem, one could focus on embedding individual subsets of the ecosystem, e.g., embeddings of TPD networks that provide the same service, or belong to the same country or continent (cf. Appendix~\ref{appendix}). The data we release contains additional information that could facilitate such analysis (see Appendix~\ref{appendix}). 

Finally, we note that here we considered undirected and unweighted networks. More generally, one can define directed and weighted TPD networks, where the link direction indicates the direction of interaction between TPDs, while the link weight in each direction may indicate the number of times the TPDs interacted in that direction. Such networks can be also constructed from the data we collected and release to the public~\cite{TPD_repo}. However, research on embedding directed and weighted networks into hyperbolic spaces is still in its infancy, and there are currently no established methods for this purpose, although some attempts have been recently made separately for directed~\cite{Zongning2020} and weighted~\cite{weighted2021} networks.

\section*{Acknowledgements}
\noindent This work is supported by the TV-HGGs project (OPPORTUNITY/0916/ERCCoG/0003) co-funded by the European Regional Development Fund and the Republic of Cyprus through the Research and Innovation Foundation.

\bibliographystyle{ACM-Reference-Format}
\bibliography{reference}


\appendix
\section{Appendix} 
\label{appendix}

Here, we provide a summary of the data that we release to the public along with related code~\cite{TPD_repo}.

\subsection{Raw data}
\label{app:raw}

We release the raw data collected in Sect.~\ref{subsec:data_collection}. This data includes all interactions between TPDs that we tracked using our browser extension from October 2017 to March 2020. The data is in the form of a csv file. Each line in the file corresponds to an HTTP interaction between two TPDs. Overall, the data contains more than 3.4M entries. The data also provides other information not used in this paper. In summary, the data contains the following information:
\begin{itemize}[leftmargin=0.4cm]
    \item \textbf{Timestamp:} the UNIX timestamp of each collected interaction between two TPDs.
    \item \textbf{FirstPartyDomain:} the URL of the visiting (first-party) domain from where each interaction between two TPDs has been triggered.
    \item \textbf{Country:} the country from where our browser extension tracked each interaction between two TPDs.
    \item \textbf{ReferrerDomain:} the TPDs that initiated each observed HTTP request.
    \item \textbf{RequestedDomain:} the URL that each observed HTTP request is directed to.
    \item \textbf{RequestType:} the resource type of each HTTP request, e.g., font, image, media, etc.
    \item \textbf{ServerIP:} the IP address of the server that responded to each HTTP request.
\end{itemize}

\subsection{FQDN list}
\label{app:FQDN_list}

The fully qualified domain name (FQDN) list contains the mapping of each FQDN to its corresponding TLD+1 domain. The mapping is performed as described in Sect.~\ref{subsec:data_processing}. The list contains 10125 entries. 

\subsection{Legal-entity list}
\label{app:legal-entities_list}

The legal-entity list is constructed as described in Sect.~\ref{subsec:data_processing}. It contains the mapping of each TLD+1 domain to its legal entity and the functionality provided by each entity (e.g., Data Management Platform, Mobile Ad Exchange, etc.). The list contains 1847 entries.

\subsection{Co-hosting list}
\label{app:co-hosting_list}

The co-hosting list is constructed as described in Sect.~\ref{subsec:data_processing}. It contains the pairs of TPDs that share the same IP address and the corresponding legal entities that the TPDs belong to. The list contains 1181 entries.


\subsection{Future-merging list} 
\label{app:chrunchbase_list}

The future-merging list is also described in Sect.~\ref{subsec:data_processing}. Each entry in the list contains a pair of TLD+1 domains and the legal entity under which they merged in the period April 2020 to November 2021. The list contains 119 entries.

As mentioned in Sect.~\ref{subsec:data_processing}, we collected the above data from the crunchbase website~\cite{crunchbase}. We provide the tool that we developed and used to automate the data collection process from the above website at~\cite{TPD_repo}.


\subsection{TLD+1 and legal-entity networks and their hyperbolic embeddings}

We also make available the edge lists of the TLD+1 and legal-entity networks constructed in Sect.~\ref{sec:graph}, as well as their corresponding hyperbolic embeddings inferred by Mercator~\cite{Mercator_repo}.

\subsection{Hyperbolic visualization tool} 
\label{app:visual_map}

Finally, we release the custom tool that we developed to visualize and interact with the hyperbolic map of the legal-entity network. The tool is based on standard web technologies (JavaScript, HTML and CSS) and can be run on any web browser. We used this tool to create  Fig.~\ref{fig:hypermap}. The tool provides the following functionalities:
\begin{enumerate}[leftmargin=0.6cm]
    \item zoom in and out on regions of interest;
    \item drag the map;
    \item select / hover over nodes to view (a) the node name (legal entity), (b) the node degree, and (c) get highlighted links between the selected node and its one-hop neighbors;  
    \item export the map as an image in different formats (svg, png and jpeg).
\end{enumerate}
 

\section{TLD+1 Networks per continent} 
\label{appendix_continent}

Here we construct the TLD+1 networks corresponding to five continents: Europe, South America, Asia, Africa and North America. To construct each network, we first map the users from which we collected data to continents, according to their country. Then, we construct each TLD+1 network by considering only the data collected from the users of the corresponding continent, following the same procedure as in Sect.~\ref{sec:graph} for the global TLD+1 network.

\begin{figure*}[ht]
\centering
    \hfill
    \begin{subfigure}[b]{0.33\textwidth}
        \centering
        \includegraphics[width=\textwidth]{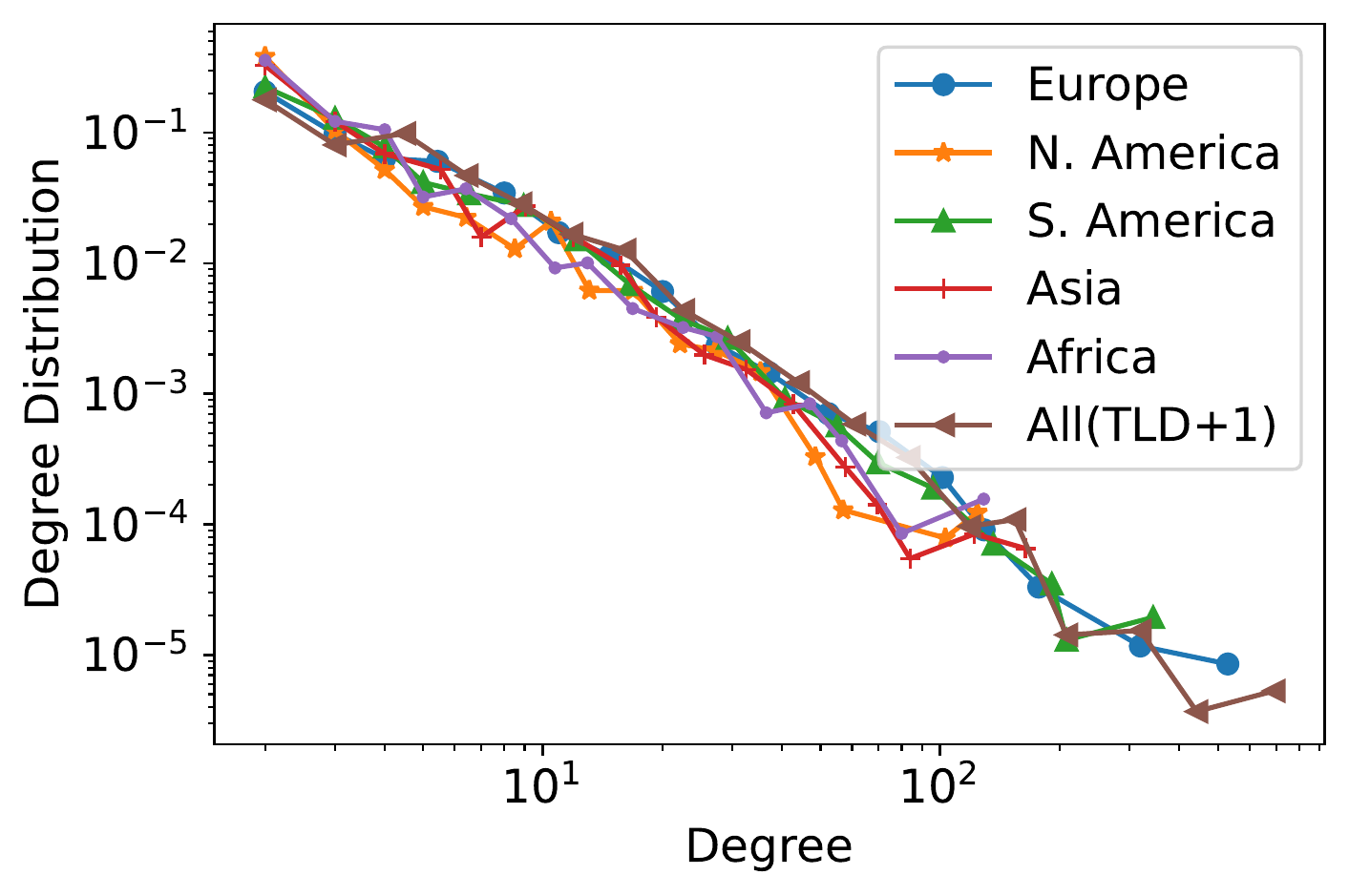}
        \vspace{-7mm}
        \caption{Degree distribution $P(k)$}
        \label{subfig:degree_distributions_cont}
    \end{subfigure}
    \hfill
    \begin{subfigure}[b]{0.33\textwidth}
        \centering
        \includegraphics[width=\textwidth]{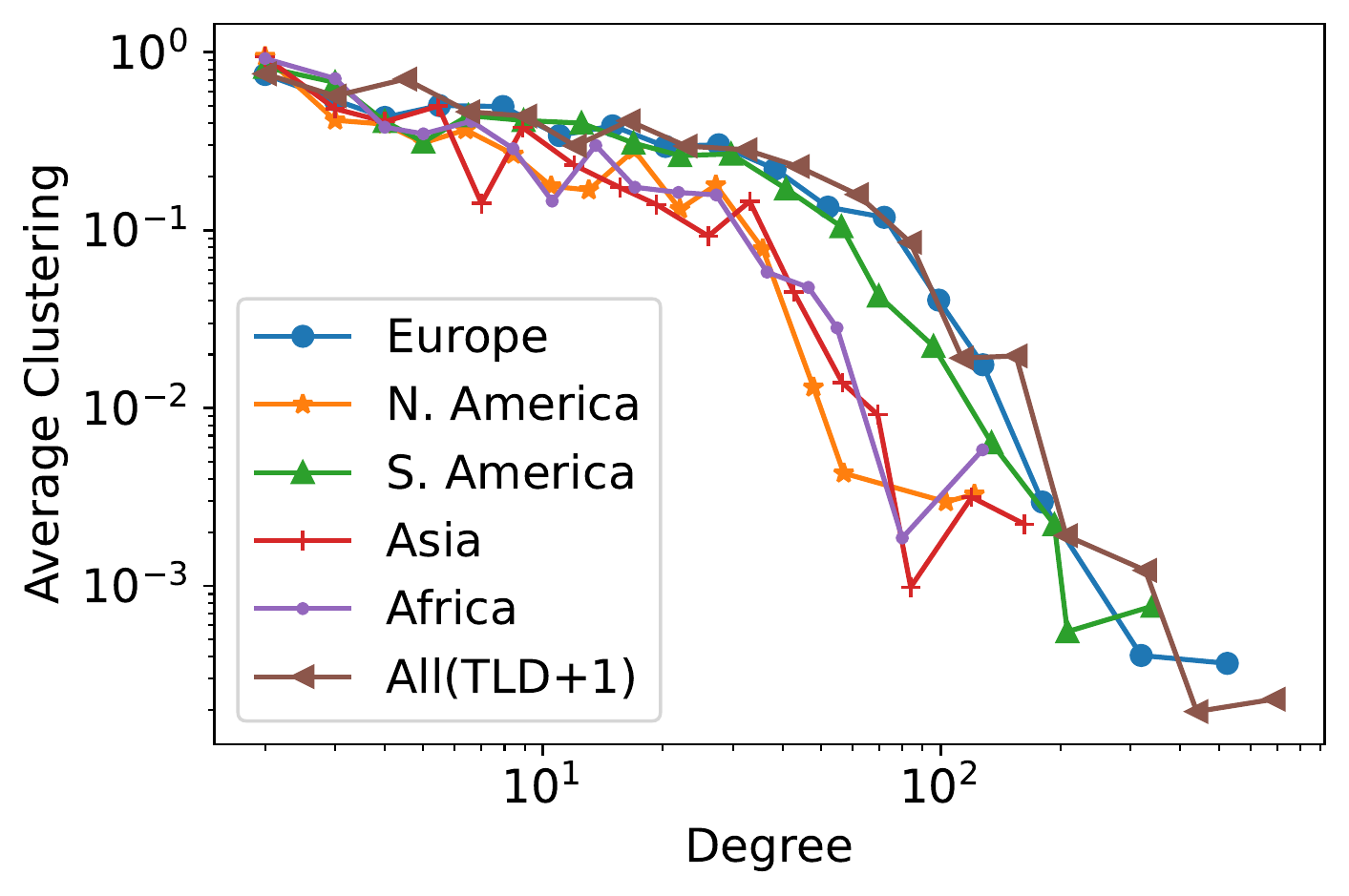}
        \vspace{-7mm}
        \caption{Average clustering $\bar{c}(k)$}
        \label{subfig:average-clustering-cont}
    \end{subfigure}
    \hfill
    \begin{subfigure}[b]{0.33\textwidth}
        \centering
        \includegraphics[width=\textwidth]{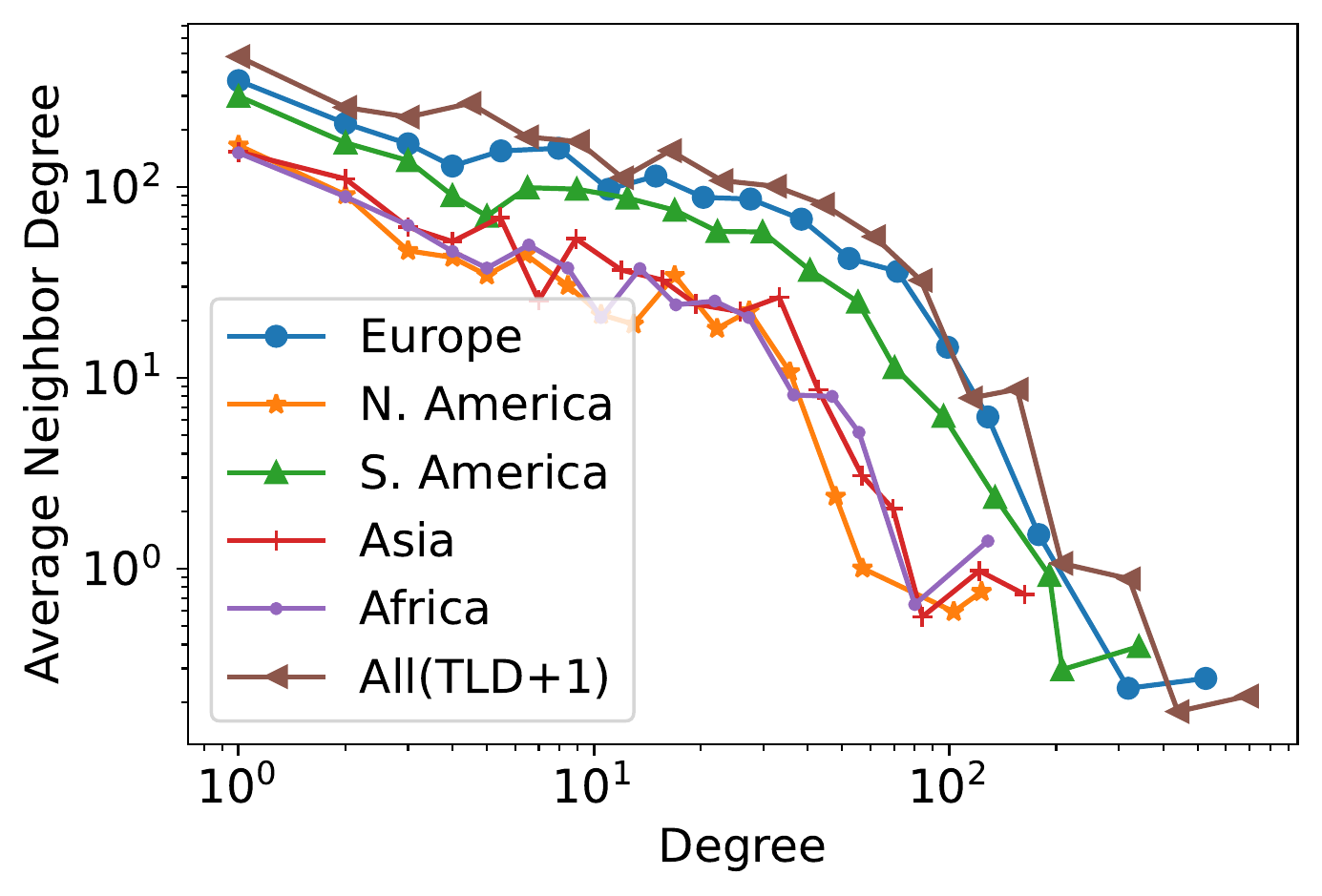}
        \vspace{-7mm}
        \caption{Average neighbor degree $\bar{k}_{\text{nn}}(k)$}
        \label{subfig:average-neighbor-degree-cont}
    \end{subfigure}
    \begin{subfigure}[b]{0.31\textwidth}
        \centering
        \includegraphics[width=\textwidth]{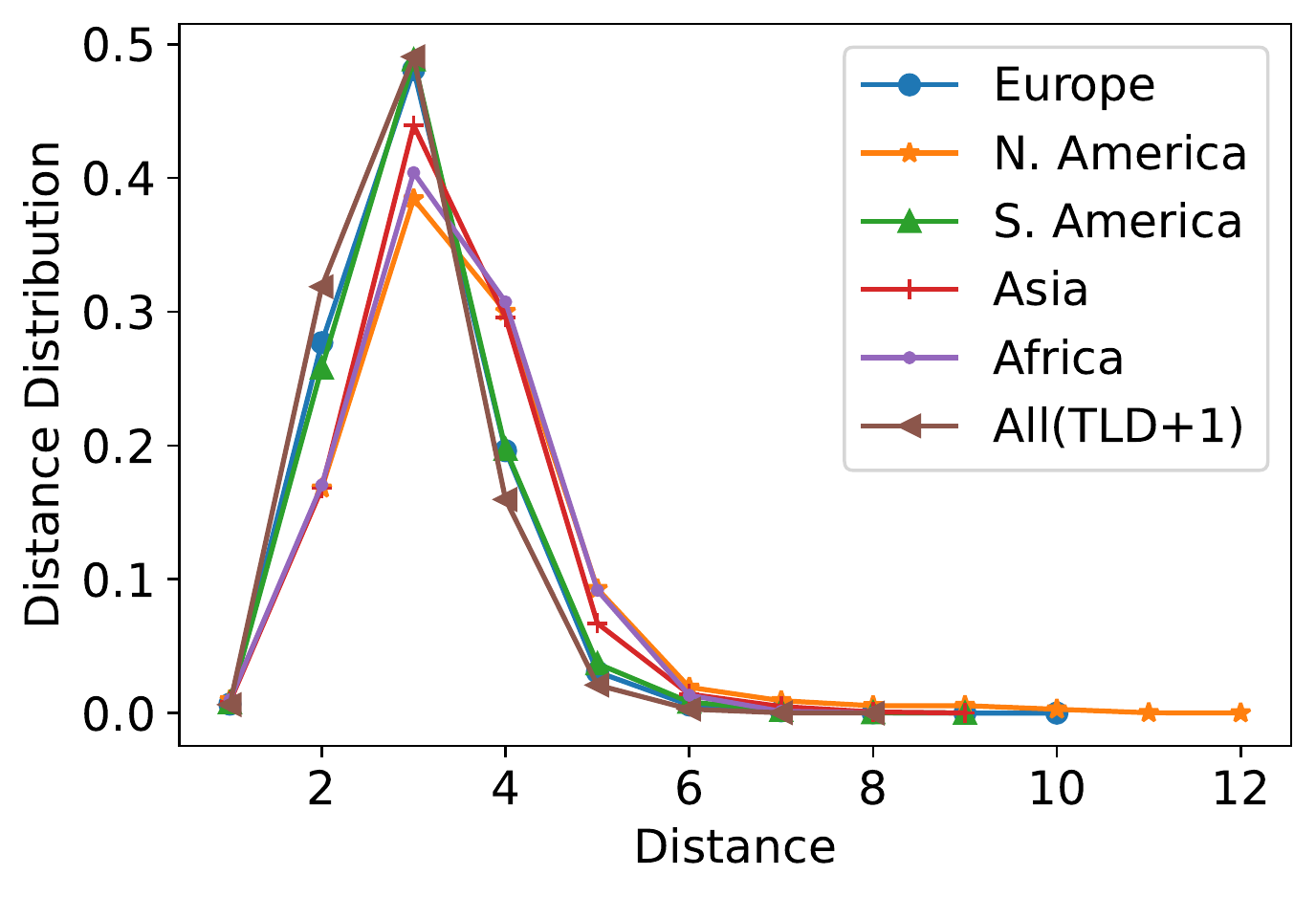}
        \vspace{-7mm}
        \caption{Distance distribution $d(l)$}
        \label{subfig:distance-distribution-cont}
    \end{subfigure}
    \begin{subfigure}[b]{0.31\textwidth}
        \centering
        \includegraphics[width=\textwidth]{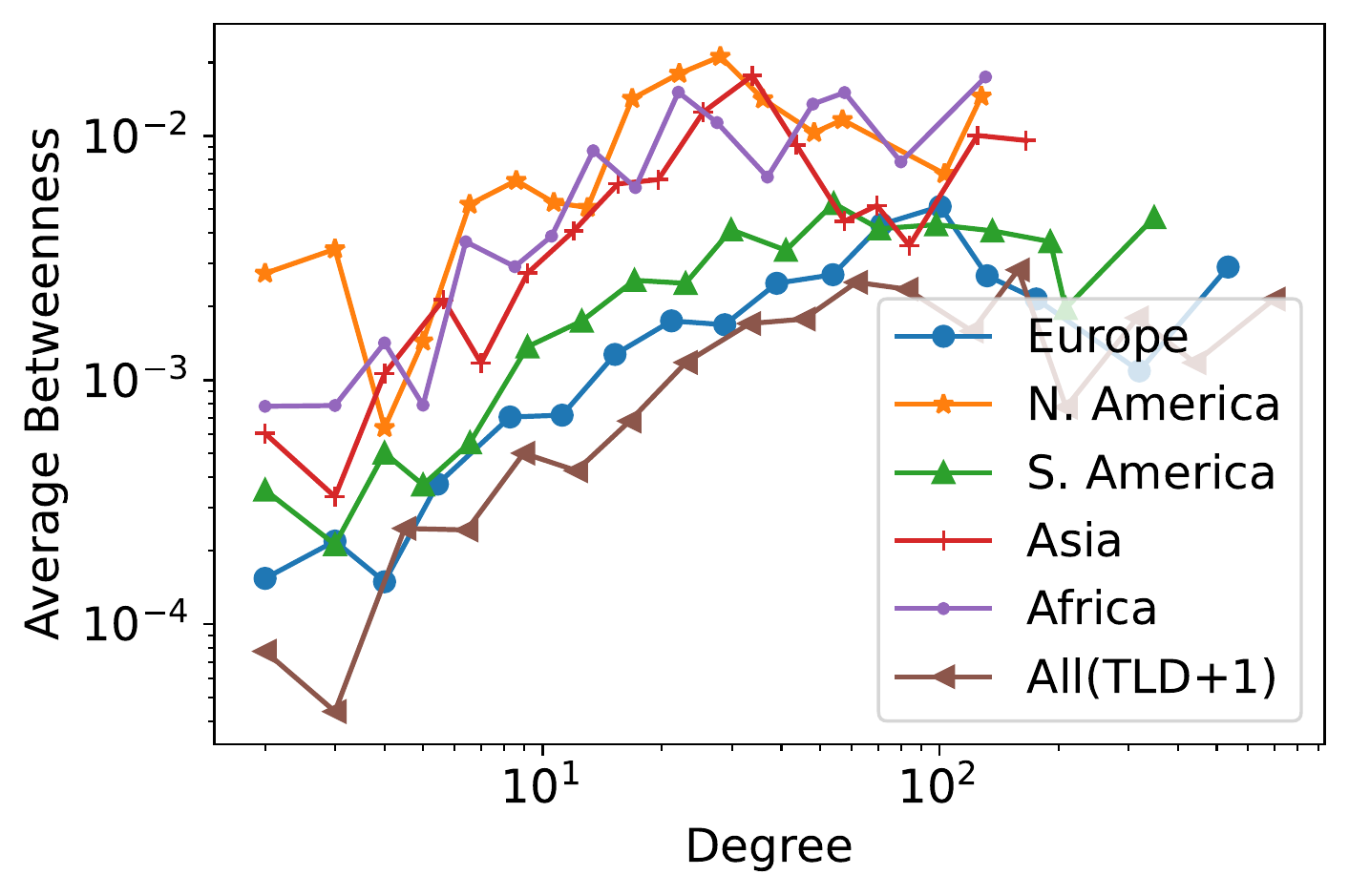}
        \vspace{-7mm}
        \caption{Average betweenness $\bar{B}(k)$}
        \label{subfig:average-betweenness-cont}
    \end{subfigure}
    \vspace{-3mm}
    \caption{Same as in Fig.~\ref{fig:structural-properties} in the main text, but for the TLD+1 networks of each individual continent. For comparison, the figure also shows the corresponding results of the global TLD+1 network considered in Fig.~\ref{fig:structural-properties}.}
    \label{fig:structural-properties-cont}
\end{figure*}

Figure~\ref{fig:structural-properties-cont} shows that the topological characteristics of the constructed TLD+1 networks are qualitatively very similar to each other and to the global TLD+1 network considered in the main text. The average clustering coefficient $\bar{c}$ in the individual TLD+1 networks is significant, ranging between $0.32$ and $0.44$, while the power-law degree distribution exponent in all cases is $\gamma \approx 2.3$ (as in the global TLD+1 network). These results suggest that the individual TLD+1 networks can be also meaningfully embedded into the hyperbolic space~\cite{Krioukov2010, Boguna2010}. However, performing and exploring such embeddings is beyond the scope of the present paper.

\begin{figure}[ht]
\centering
\includegraphics[width=0.85\columnwidth]{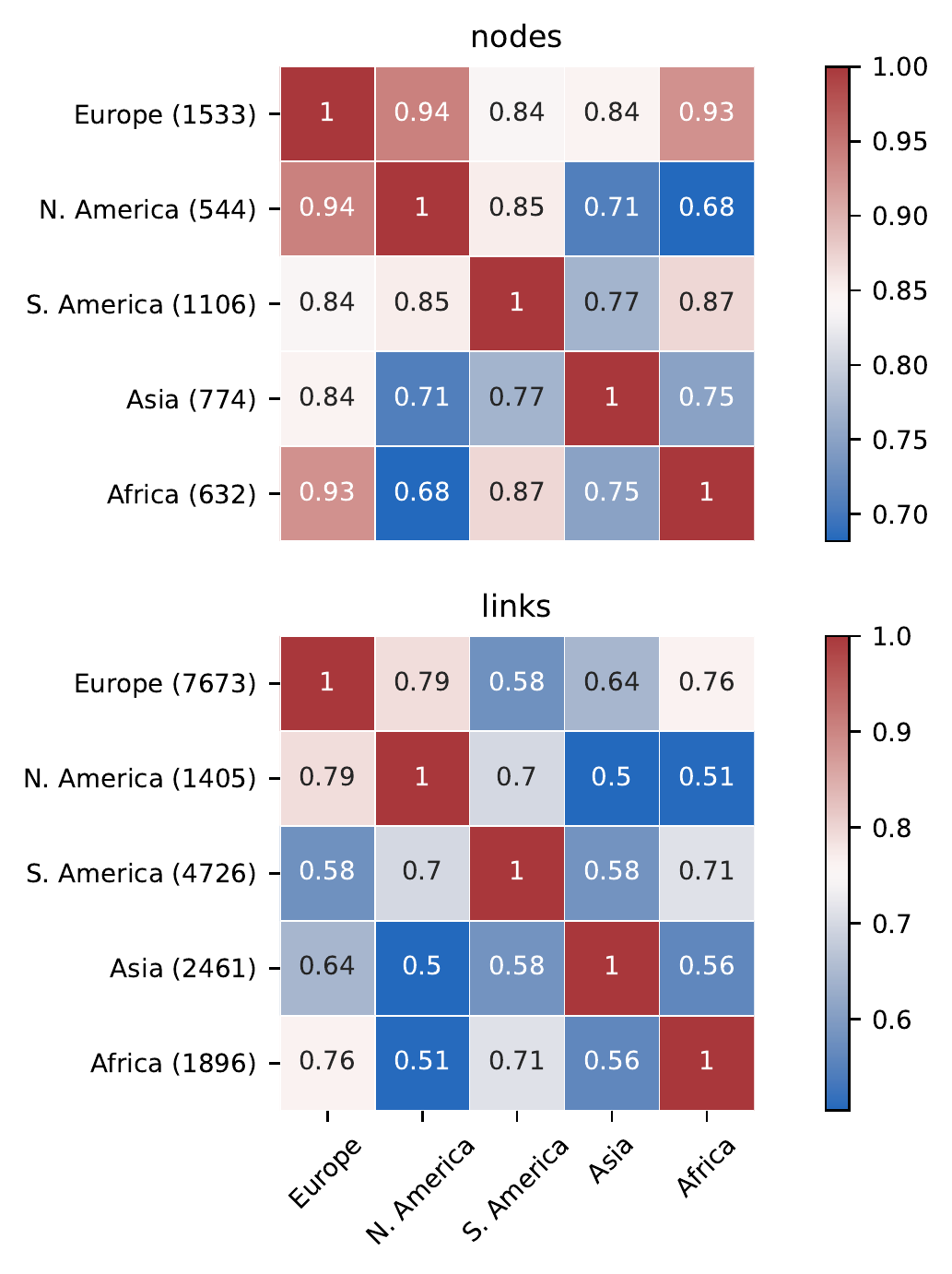}
\vspace{-4mm}
\caption{The percentage of overlapping nodes (top) and links (bottom) among the continents' TLD+1 networks. The numbers in parentheses indicate the number of nodes and links in each network.}
\label{fig:continents_overlap}
\end{figure}

Finally, Fig.~\ref{fig:continents_overlap} shows that the individual TLD+1 networks exhibit high percentages of overlapping nodes and links.

\end{document}